\documentclass[12pt,preprint]{aastex}
\usepackage{natbib}

\bibpunct{(}{)}{;}{a}{}{,}

\newcommand{\ginga}{\it Ginga}
\newcommand{\xte}{\it RXTE}
\newcommand{\integ}{\it INTEGRAL}
\newcommand{\swift}{\it Swift}

\newcommand{\ea}{$E_{\rm a1}$}
\newcommand{\eb}{$E_{\rm a2}$}
\newcommand{\ec}{$E_{\rm a3}$}

\newcommand{\ch}{$\chi_{\nu}^{2}$}
\newcommand{\ergs}{erg s$^{-1}$}

\slugcomment{nothing}

\shorttitle{Cyclotron energy variation in X~0331+53}
\shortauthors{Nakajima et al.}

\begin{document}

\title{Energy-Dependent Harmonic Ratios of the Cyclotron Features of X0331+53 in the 2004-2005 Outburst}

\author{M. Nakajima\altaffilmark{1,2}, T. Mihara\altaffilmark{2} and K. Makishima\altaffilmark{2,3}}
\altaffiltext{1}{School of Dentistry at Matsudo, Nihon University, 2-870-1, Sakae-cho Nishi, Matsudo-City, Chiba, JAPAN 271-8587}
\altaffiltext{2}{Cosmic Radiation, The Institute of Physical and Chemical Research (RIKEN), 2-1 Hirosawa, Wako, Saitama, JAPAN 351-0198}
\altaffiltext{3}{Department of Physics, University of Tokyo, 7-3-1 Hongo, Bunkyo-ku, Tokyo, JAPAN 113-0033}

\begin{abstract}

We report on changes of the cyclotron resonance energies of 
the recurrent transient pulsar, X0331+53 (V0332+53).
The whole {\xte} data acquired in the 2004-2005 outburst were utilized. 
The $3-80$ keV source luminosity varied between $1.7\times10^{36}$ 
and $3.5\times10^{38}$ {\ergs}, assuming a distance of 7 kpc. 
We confirmed that the fundamental cyclotron resonance energy changed 
from $\sim22$ to $\sim27$ keV in a clear anti-correlation to 
the source luminosity, and without any hysteresis effects 
between the rising and declining phases of the outburst. 
In contrast, the second harmonic energy changed from $\sim49$ to $\sim54$ keV, 
implying a weaker fractional change as a function of the luminosity. 
As a result, the observed resonance energy ratio between 
the second harmonic and the fundamental 
was $\sim2.2$ when the source was most luminous, 
whereas the ratio decreased to the nominal value of 2.0 
at the least luminous state. 
Although the significance of this effect is model dependent,
these results suggest that the fundamental and second harmonic resonances 
represent different heights in the accretion column, 
depending on the mass accretion rate.

\end{abstract}


\section{INTRODUCTION}
\label{sec1}

Magnetic field strengths on the surface of accreting X-ray pulsars 
can be measured accurately by observing 
Cyclotron Resonant Scattering Feature (CRSF), 
because the fundamental electron cyclotron resonance energy 
{\ea} is described as 
$E_{a1} = 11.6 \ B_{12} (1+z_g)^{-1}\ ({\rm keV})$. 
Here, $B_{12}$ is the magnetic field strength in units of $10^{12}$ Gauss, 
and $z_g$ is the gravitational redshift. 
Applying this relation to X-ray detected CRSFs, 
the magnetic field strengths of $\sim$15 X-ray pulsars 
have been accurately measured 
\citep[and references therein]{tru78,whi83,tm95,max99,cob02}. 
The results are clustered in a relatively narrow range of 
$(1-4)\times10^{12}$ G \citep{max99}.

Since the cyclotron resonant scattering occurs near the pulsar surface, 
the resonance energies were believed to be constant and 
intrinsic to each pulsar. 
However, an unexpected CRSF energy change was found from 
the recurrent transient pulsar 4U~0115+63 with the {\it Ginga} observations 
performed in 1990 and 1991 
\citep{tm95,tm98,tm04}; 
in the 1991 minor outburst when the luminosity was $\sim\frac{1}{7}$ 
of that in the 1990 and other typical outbursts of this object, 
the value of {\ea} was $\sim1.4$ times higher than 
the so far reported {\ea}$\sim11$ keV. 
In order to interpret the change in {\ea}, 
Mihara et al.\ (2004) proposed that 
the height of cyclotron scattering region in the accretion column changes 
depending on the X-ray luminosity. 
A further study of the luminosity-dependent CRSFs change 
was conducted by Nakajima (2006), Nakajima et al.\ (2006a,b), 
and Tsygankov et al.\ (2007), 
using {\it Rossi X-ray Timing Explorer} ({\xte}) data of 
4U~0115+63 acquired in 1999 and 2004. 
According to these results, 
{\ea} increased from $\sim10$ to $\sim16$ keV 
as the source luminosity decreased across a narrow range of 
$(2-4)\times10^{37}$ {\ergs} 
(at an assumed distance of 7 kpc; Negueruela \& Okazaki 2001). 
In addition, the second harmonics observed at $\sim20$ keV 
disappeared as {\ea} started to change.

This luminosity-dependent change in the CRSF energy 
has been found from another source, 
X0331+53 (V0332+53), with {\it Ginga}, {\integ} and {\xte} observations 
(Mihara et al.\ 1998; Mowlavi et al.\ 2006; Nakajima 2006; 
Tsygankov et al.\ 2006, 2009). 
This source is a recurrent transient pulsar, and 
exhibits a very prominent fundamental CRSF at $\sim28$ keV of 
the X-ray spectrum, 
as first discovered with {\ginga} \citep{max90a} 
after a suggestion with {\it Temma} \citep{max90b}. 
Furthermore, its two higher harmonics have been discovered 
at energies of $\sim50$ and $\sim75$ keV 
\citep{cob05,kre05,pot05}. 
Thus, X0331+53 is the second object 
which has three or more CRSFs in the spectrum, following 4U~0115+63 
\citep{san99,hei99}. 
However, compared with the results from 4U~0115+63, 
this object exhibited two intriguing differences 
(Mowlavi et al.\ 2006; Nakajima 2006; Tsygankov et al.\ 2006, 2009). 
One is that the change in {\ea} started at a higher luminosity, 
$1\times10^{38}$ {\ergs}, 
at an assumed distance of 7 kpc \citep{neg99}. 
The other is that the second harmonic absorption feature persisted in 
the X-ray spectra through an outburst.

Including the above cases of 4U~0115+63 and X0331+53, 
the behavior of the fundamental CRSF 
has been extensively studied so far. 
In contrast, details of the second harmonic, 
including its possible luminosity dependence, 
are less understood, 
primarily because of lower data quality at higher energies. 
With this in mind, 
we analyzed the whole {\xte} (ASM, PCA and HEXTE) data of X0331+53, 
acquired in a 2004-2005 outburst. 
As a result, we have discovered that 
the second CRSF energy {\eb} possibly depends 
more weakly on the luminosity than {\ea}. 
The {\eb}/{\ea} ratio was $\sim2.2$ when the source was luminous, 
while it approached the nominal value of 2.0 toward lower luminosities. 
These results suggest that 
the two resonances take place at different heights 
in the same accretion columns, 
and that {\eb}, formed at a lower height, 
provides a more reliable estimate of 
the surface field strength.

\section{OBSERVATIONS and DATA REDUCTION}
\label{sec2}

In order to study the luminosity-dependent changes of 
{\ea} and {\eb} of X0331+53, 
here we utilized all the {\xte} data sets of this transient pulsar 
acquired in the 2004-2005 outburst. 
Figure \ref{f1}a shows the whole light curve of this outburst 
acquired with the All Sky Monitor (ASM; Levine et al.\ 1996) onboard {\xte}. 
An abrupt brightening was detected on 2004 November 25 \citep{swa04}, 
and the $2-12$ keV flux continued to increase up to 1.1 Crab \citep{rem04}. 
As revealed by the ASM, the X-ray intensity declined toward 
middle of 2005 February, and exhibited a small recovery at 
the beginning of 2005 March, presumably synchronized with 
the 34.25 day orbital period \citep[and references therein]{zha05}. 
At the outburst peak, a number of 
target-of-opportunity observations were 
performed by {\integ}, {\swift}, and {\xte}.

From 2004 November 27 through 2005 March 27, 
108 pointing observations were made 
with the Proportional Counter Array (PCA; Jahoda et al.\ 2006) and 
the High Energy X-ray Timing Experiment (HEXTE; Rothschild et al.\ 1998) 
on board {\xte}.
Based on the inspection of the operation status of 
the five proportional counter units (PCU0$\sim$PCU4), 
we have selected the data sets acquired with PCU2, 
which worked throughout this outburst. 
As reported by Pottschmidt et al.\ (2005), 
HEXTE cluster A did not function from 2004 December 13 to 2005 January 14. 
Thus, we utilize only the data acquired with HEXTE cluster B.

Using part of the {\xte} data sets that 
were acquired in the descent phase, 
the behavior of the fundamental CRSF was already studied 
\citep{mow06,m06,tsy06}. 
In the present paper, 
we complement these studies of the descent phase, 
by analyzing those datasets which were left unused. 
In addition, we analyze the ascent-phase datasets to examine 
whether the CRSF energy exhibits any hysteresis effects similar to those 
suggested by the 4U~0115+63 data \citep{m06a}.

During the 4 months of observation, 
several energetic background events, 
such as large solar flares and 
precipitations of high-energy particles trapped in the radiation belts, 
affected some of the data sets. 
In order to exclude such low-quality data, 
we inspected the electron-rate light curves, and 
have selected 86 data sets with good quality as 
listed in Table \ref{t1}. 
To analyze them, we utilized the HEAsoft version 6.0.5. 
We assume 1\% systematic error for all energy bins 
of the PCA data (same as in Nakajima et al.\ 2006a). 
All of the errors presented in this paper are 90\% confidence levels.

\section{DATA ANALYSIS AND RESULTS}
\label{sec3}

\subsection{Analysis of Representative Spectra}
\label{sec3.1}

\begin{figure}
\epsscale{0.7}
\plotone{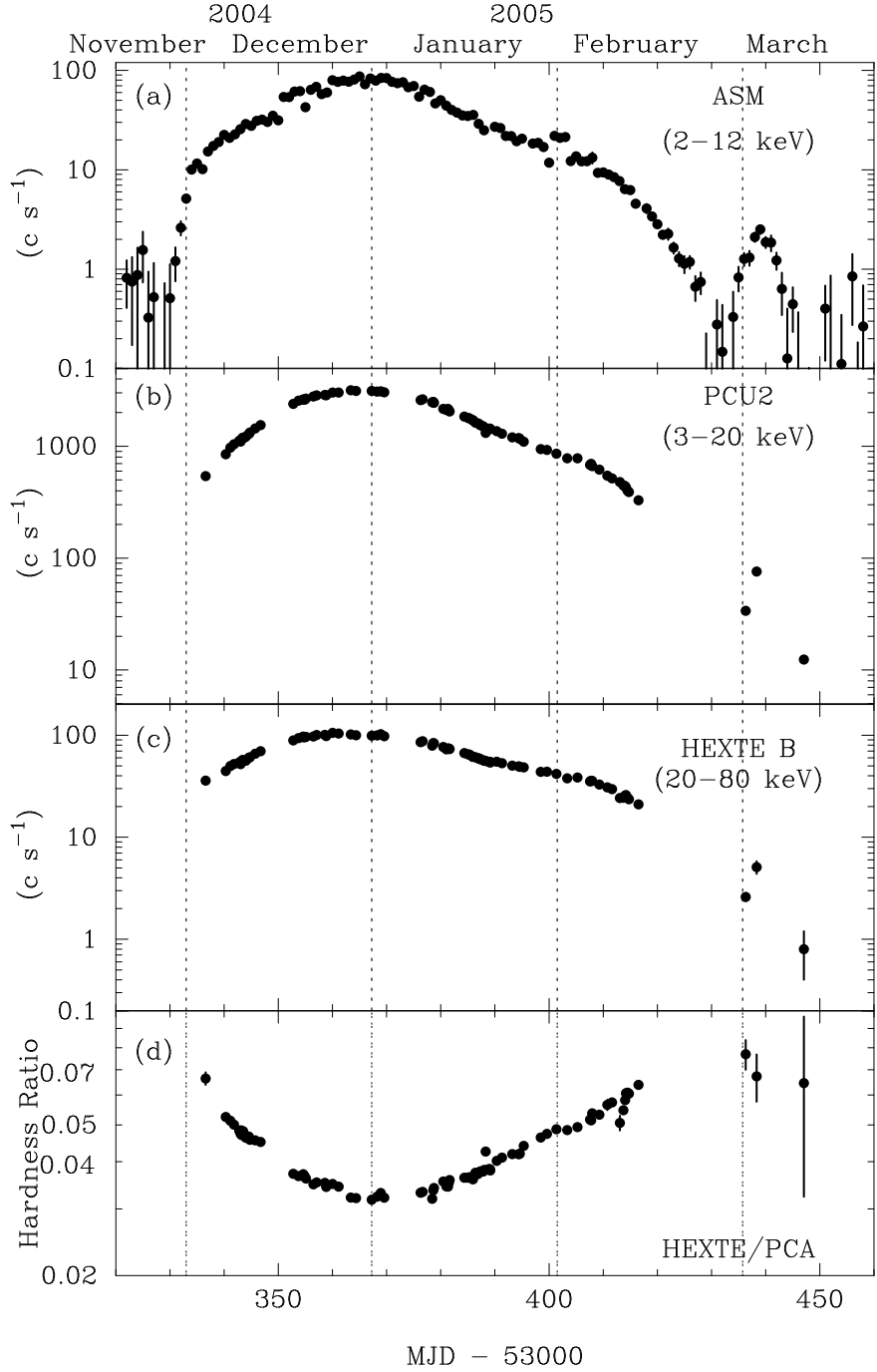}
\caption{Background-subtracted count rates and hardness ratios of X0331+53 obtained from the {\xte} observations. Dashed lines indicate the days of periastron passage (Zhang et al.\ 2005, and references therein).  (a) The $2-10$ keV ASM lightcurve, with 1 day binning.  (b) The $3-20$ keV PCU2 lightcurve.  (c) The $20-80$ keV HEXTE cluster B lightcurve.  (d) The hardness ratio, defined as the ratio between the HEXTE and PCU count rates. }
\label{f1}
\end{figure}

As shown in Figure \ref{f1},  
the intensity of X0331+53 changed largely 
in the ASM, PCA and HEXTE energy bands 
during this outburst. 
In addition, the hardness-ratio exhibited 
an anti-correlation to the source count rates 
as shown in Figure \ref{f1}d. 
In order to investigate the luminosity related CRSFs changes 
through this outburst, 
we need to establish a unified spectral modeling 
which can be applied consistently to the data sets at all luminosity levels. 
So, we first examined 
pulse-phase-averaged PCA and HEXTE spectra 
from several representative data sets.

\begin{figure}
\epsscale{1.0}
\plotone{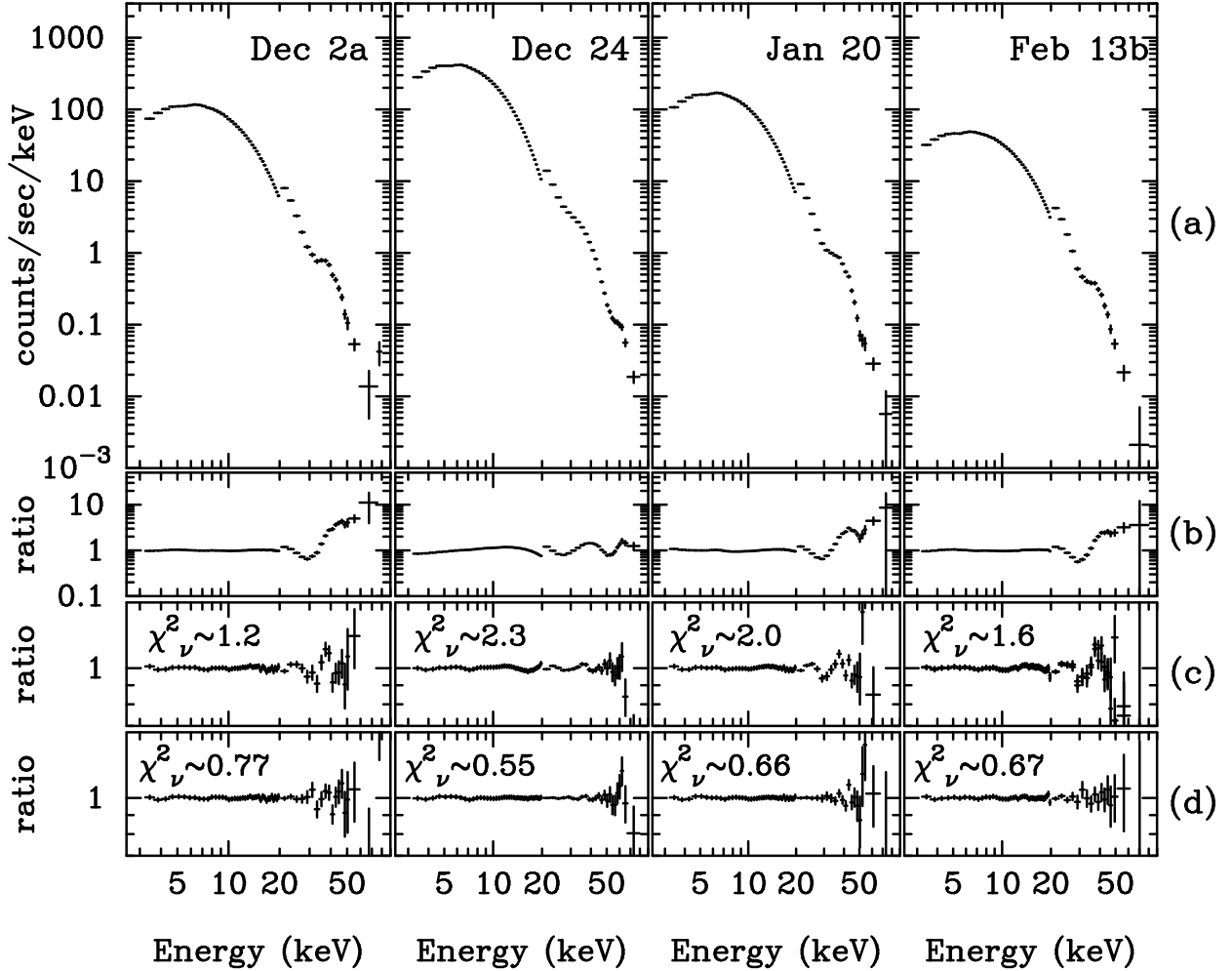}
\caption{Representative pulse-phase-averaged spectra of X0331+53, obtained on Dec 2a, Dec 24, Jan 20, and Feb 13b.  (a) The background-subtracted and response-inclusive PCA ($3-20$ keV) and HEXTE ($20-80$ keV) spectra.  (b) The PCA and HEXTE spectra in panel (a), normalized to the best-fit NPEX models which incorporates no absorption factors.  (c) The data to NPEX$\times$CYAB2 model ratios.  (d) The data to NPEX$\times$CYAB2$\times$GABS model ratios. Detailed model parameters are given in Table \ref{t2}.  }
\label{f2}
\end{figure}

Since many of the data sets have rather short exposure, 
we selected four representative ones sets 
with relatively long exposures, 
denoted in Table \ref{t1} 
as Dec 2a, Dec 24, Jan 20, and Feb 13b; 
these represent the beginning, the peak, the declining phase, 
and the end of the outburst, respectively. 
Figure \ref{f2}a shows the background-subtracted 
PCU2 ($3-20$ keV) and HEXTE B ($20-80$ keV) spectra 
obtained on these four occasions. 
The PCA background spectra were all estimated 
with the bright-background model \citep{jah06}, 
while the HEXTE backgrounds were extracted from 
the off-source position which is offset from the source 
by 1.5 degrees \citep{rot98}. 
The four spectra, even in their raw forms, clearly reveal 
the prominent fundamental CRSF at $\sim30$ keV, 
which was observed in the previous (Makishima et al.\ 1990a,b) and 
the present 
(Kreykenbohm et al.\ 2005; Pottschmidt et al.\ 2005; Mowlavi et al.\ 2006; 
Nakajima 2006; Tsygankov et al.\ 2006, 2009) 
outburst of this object. 
Furthermore, the brighter two spectra clearly show 
the second harmonic resonance at $\sim60$ keV.

From early days (e.g. White et al.\ 1983), 
continuum spectra of accretion powered pulsars were 
approximated by a power-law modified by exponential cutoff. 
In this paper, we employ
NPEX (Negative and Positive power-law with EXponential) model 
\citep{tm95,max99}, which is described as 
\begin{equation}
N(E) = ( A_1 E^{-\alpha_1} + A_2 E^{+2.0} ) ~ \exp \left( -\frac{E}{kT} \right),
\label{e1}
\end{equation}
where $E$ is the X-ray energy in units of keV, 
$A_1$ and $\alpha_1$ are the normalization and photon index of the negative 
power-law, respectively, $A_2$ is the normalization of the positive power-law, 
and $kT$ represents the cutoff-energy in units of keV. 
This model successfully reproduced the continuum spectra
of 4U~0115+63 (Mihara et al.\ 2004; Nakajima 2006; Nakajima et al.\ 2006ab) 
at various luminosity levels.
As already reported by several authors
(e.g. Makishima et al.\ 1999; Coburn et al.\ 2002),
the determinations of the cyclotron line parameters are affected 
by the continuum and the line modeling.
Employing this model, we can compare the results of this work 
with the previous ones \citep{m06a}.
Therefore, we have selected this model.

We first attempted to fit the 4 spectra with the NPEX model, 
with $A_1$, $A_2$, $\alpha_1$, and $kT$ all left free. 
Since the PCA background estimations are only good to a few percent, 
we further adjusted the background normalizations 
down to a level of $\sim$1\%, 
following the method described in the {\it RXTE cook book} 
\footnote{See http://heasarc.gsfc.nasa.gov/docs/xte/recipes/cook\_book.html}. 
However, as expected, the NPEX model 
gave acceptable fits to none of the four spectra. 
As revealed clearly by the data-to-model ratio in Figure \ref{f2}b, 
all the four spectra exhibit the fundamental CRSF strongly at $\sim30$ keV. 
The second harmonic CRSF is observed not only in the brighter two spectra, 
but also in the two fainter ones. 
In addition, the spectrum on Dec 24 exhibits 
evidence of the third harmonic CRSF at $\sim75$ keV, 
as already reported by 
Coburn et al.\ (2005), Kreykenbohm et al.\ (2005), and 
Pottschmidt et al.\ (2005). 
Below, we concentrate on the fundamental and the second harmonic. 
The fluorescent Fe K$_\alpha$ lines at 6.4 keV with 
a width of $\sigma_{{\rm Fe}} = 0.5$ keV are also taken into account 
\citep{pot05,m06,tsy06}.

In order to evaluate the fundamental and second CRSF parameters, 
we next introduce a cyclotron absorption model. 
Although some authors used Gaussian-shaped absorption cross section 
\citep{cob02,kre04,pot05,kre05,mow06,kre06,klo08}
to reproduce the CRSF, 
we employ the cyclotron absorption (CYAB) factor 
which has been used successfully 
(Clark et al.\ 1990; Makishima et al.\ 1990a, 1999; Mihara 1995; 
Mihara et al.\ 1998; Nakajima 2006; Nakajima et al.\ 2006ab; 
Tsygankov et al.\ 2006, 2007, 2009). 
The CYAB factor is described as 
\begin{equation}
C_{i}(E) = \exp{ \biggl\{ -\frac{D_{i}~(WE/E_{ai})^2}{(E-E_{ai})^2+W_i^2} \biggl\} } ~~~ (i=1,2),
\label{e2}
\end{equation}
where $E_{ai}$ is the resonance energy, $W_i$ is the resonance width, 
$D_i$ is the resonance depth, and $i$ is the harmonic number 
(with $i=1$ fundamental). 
Using high-quality {\it Suzaku} data, 
Enoto et al.\ (2008) confirmed that 
eq.\ref{e2} can successfully reproduce 
the 36 keV CRSF of the X-ray pulsar Her X-1, 
whereas the Gaussian-shaped absorption cross-section is less successful.

Since we already know that X0331+53 has the multiple CRSFs, 
we attempted to fit the data with an NPEX model 
multiplied by two CYAB factors 
(hereafter NPEX$\times$CYAB2 model). 
The values of {\ea} and {\eb} were both left to vary independently, 
rather than constrained as {\eb}$=2${\ea}. 
As a result, 
the fits were much improved compared to the NPEX fit, 
as shown by the data-to-model ratios in Figure \ref{f2}c. 
Due to the complex shape of the fundamental CRSFs, however, 
the NPEX$\times$CYAB2 model is not yet fully successful, 
with $\chi_\nu^{2}\geq 1.2$. 
This fact was already reported from previous studies 
\citep{max90a,tm95,kre05,pot05}.

In order to better describe the fundamental CRSF, 
Kreykenbohm et al.\ (2005) and Pottschmidt et al.\ (2005) 
introduced an additional Gaussian absorption (GABS) factor 
with a similar energy but a different width, 
thus constructing a nested double-Gaussian absorption profile. 
Therefore, 
let us try to explain away the fit residuals employing 
an additional GABS factor 
\footnote{The original GABS model in XSPEC v11.3.2p had a bug related to the energy binnings as reported by Kitaguchi et al.\ (2007).  In this paper, we utilized an updated version of GABS model with the bug fixed, provided by Keith Arnaud. }, 
which is described as 
\begin{equation}
G(E) = \exp{ \left[ -\tau_{{\rm ga}} ~ \exp{ \biggl\{ -\frac{1}{2} \left(
\frac{E-E_{{\rm ga}}}{\sigma_{{\rm ga}}} \right) ^2 \biggl\} } \right] },
\label{e3}
\end{equation}
where $E_{{\rm ga}}$ is the line-center energy, 
$\sigma_{{\rm ga}}$ is the line width, 
and $\tau_{{\rm ga}}$ is the optical depth at the line center. 
Although a more sophisticated CRSF modeling 
has been introduced by Sch\"{o}nherr et al.\ (2007),
we employ the CYAB and GABS models in the present work, 
so that the results can be directly compared with previous 
studies of the luminosity dependence of CRSFs 
(Makishima et al.\ 1999; Mihara et al.\ 2004; Nakajima et al.\ 2006).

\begin{figure}
\epsscale{0.8}
\plotone{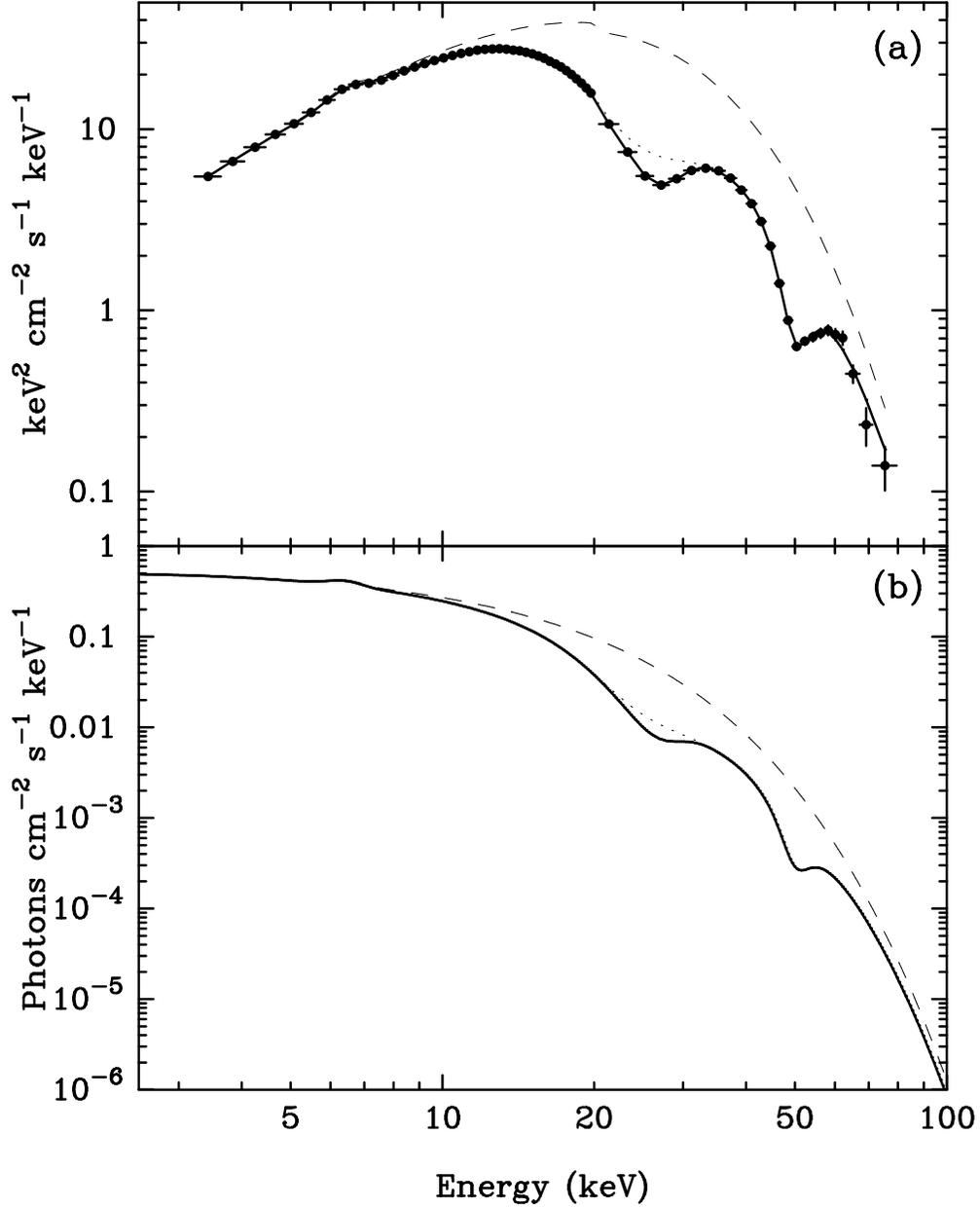}
\caption{(a) The same Dec 24 spectrum as shown in Figure \ref{f2}a, but presented in the deconvolved $\nu F_\nu$ form. The filled circles represent the data, and the solid lines are the best-fit NPEX$\times$CYAB2$\times$GABS model.  The dotted curve represent the best-fit model, from which the GABS absorption factor is removed.  The dashed lines display the NPEX continuum, obtained by further resetting the CYAB2 factor to zero.  (b) The best-fit model in its incident form. Meanings of the solid, dotted, and dashed curves are the same as in panel a. }
\label{f3}
\end{figure}

We fitted the four spectra by the NPEX$\times$CYAB2$\times$GABS model, 
namely the same NPEX$\times$CYAB2 model but 
further multiplied by eq.\ref{e3}, in which 
the three GABS parameters ($E_{{\rm ga}}$, $\sigma_{{\rm ga}}$, 
$\tau_{{\rm ga}}$) were left free 
but $E_{{\rm ga}}$ was given an initial value close to {\ea}. 
Due to low statistics in the higher energies, 
we fixed {\eb} at 2{\ea} in fitting the Dec 2a and Feb 13b data, 
while we left {\ea} and {\eb} both free and independent 
for the brighter two data sets. 
Then, as illustrated in Figure \ref{f2}d, 
the four spectra have been described successfully with this model. 
The derived best-fit parameters  
are summarized in table \ref{t2}. 
To see the configuration of the nested cyclotron models 
for the fundamental resonance, 
the spectrum observed on Dec 24 is presented in Figure \ref{f3} 
in its $\nu F_\nu$ form. 
Thus, the CYAB factor explains the gross shape of the fundamental CRSF, 
while the remaining narrow core is represented 
by the additional GABS factor. 
The value of $E_{{\rm ga}}$ agrees, within a few percent, with 
$E_{a1} \left\{ 1 + (W_1/E_{a1})^2 \right\}$, or $\sim 1.2${\ea} 
in the present case, 
where the CYAB factor becomes deepest \citep{tm95,max99}. 
For comparison, 
Kreykenbohm et al.\ (2005) and Pottschmidt et al.\ (2005) applied 
nested two GABS factors to the fundamental CRSF 
and obtained two Gaussian centroids which differs by $10-20$\%.

\begin{figure}
\epsscale{1.2}
\plottwo{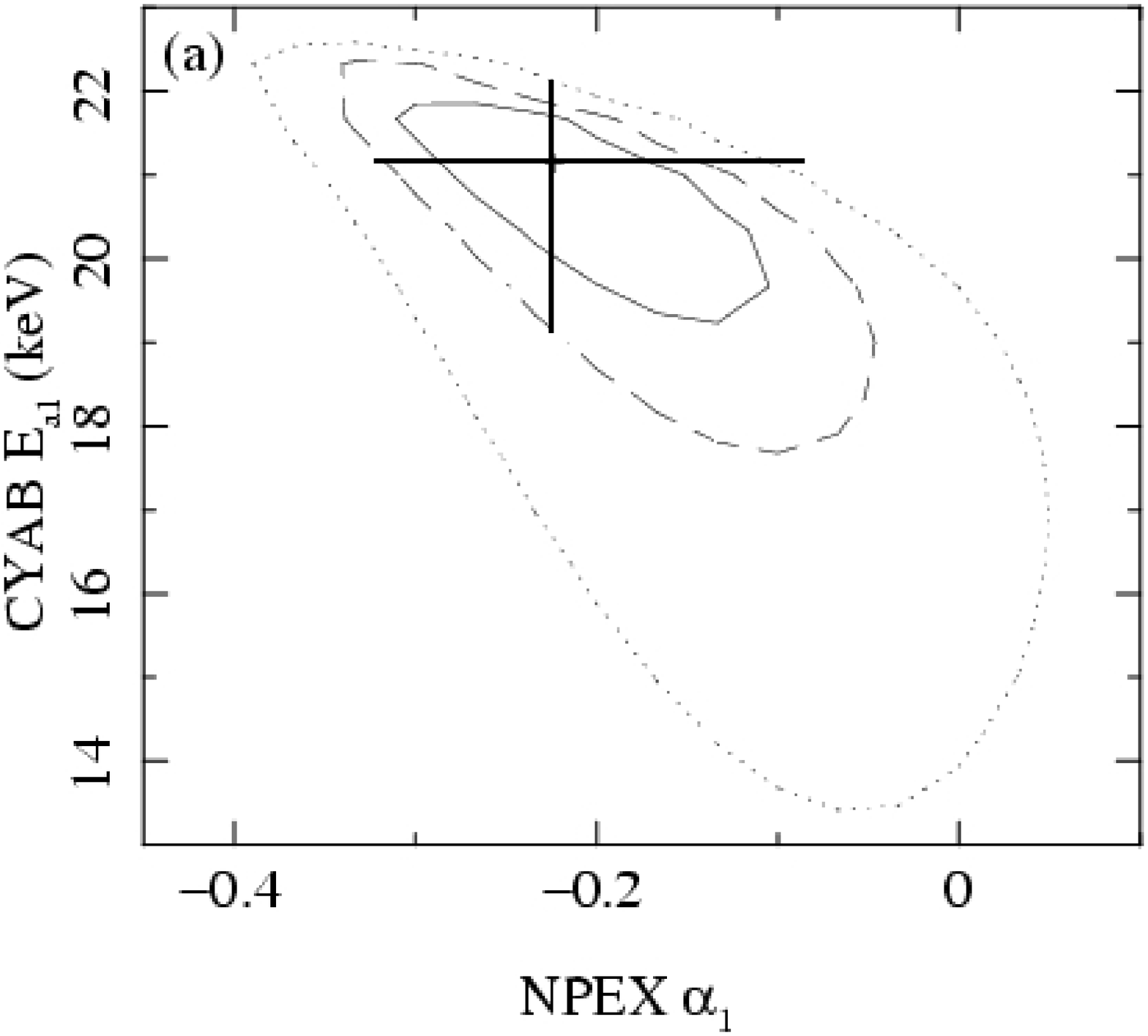}{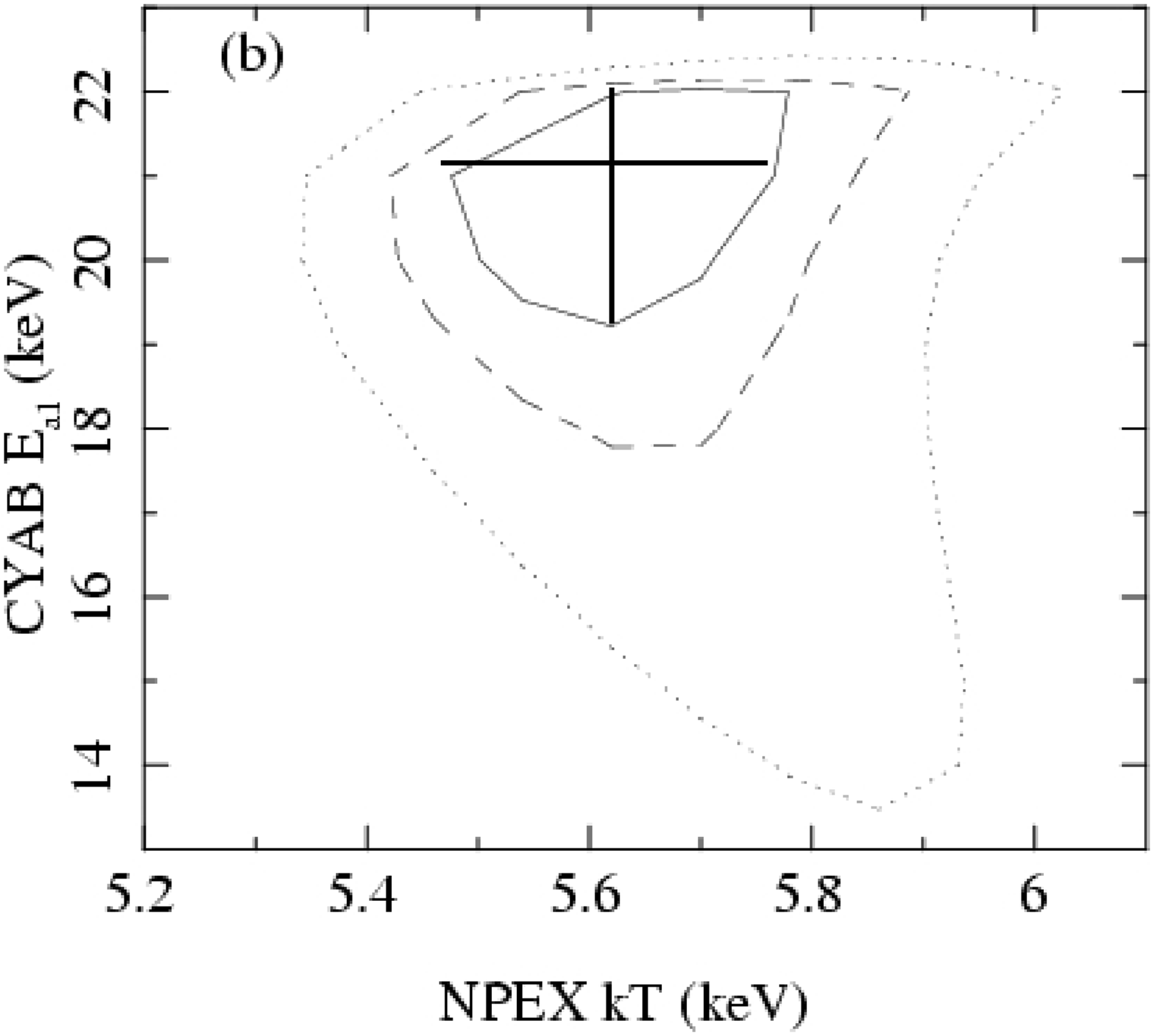}

\plottwo{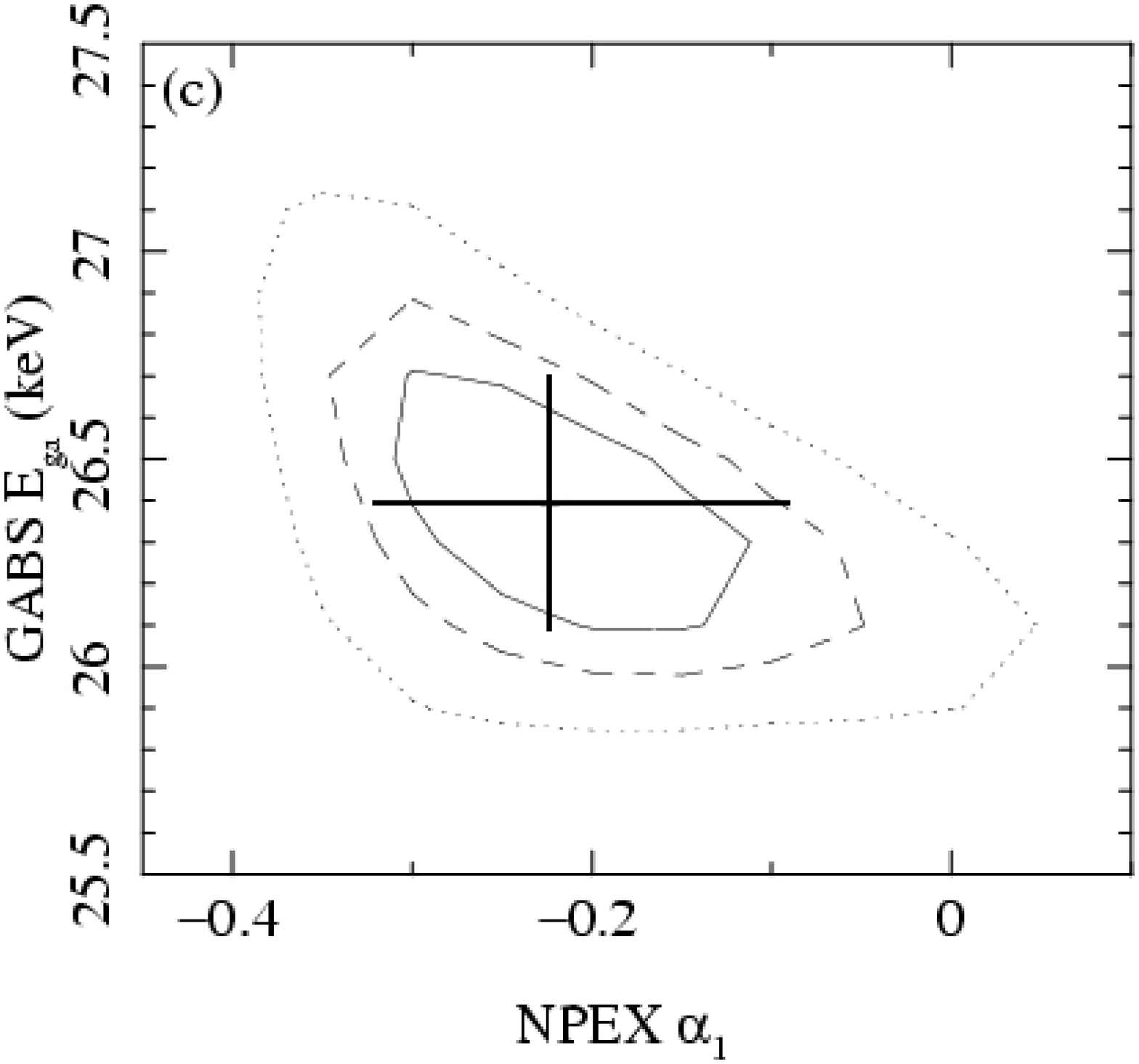}{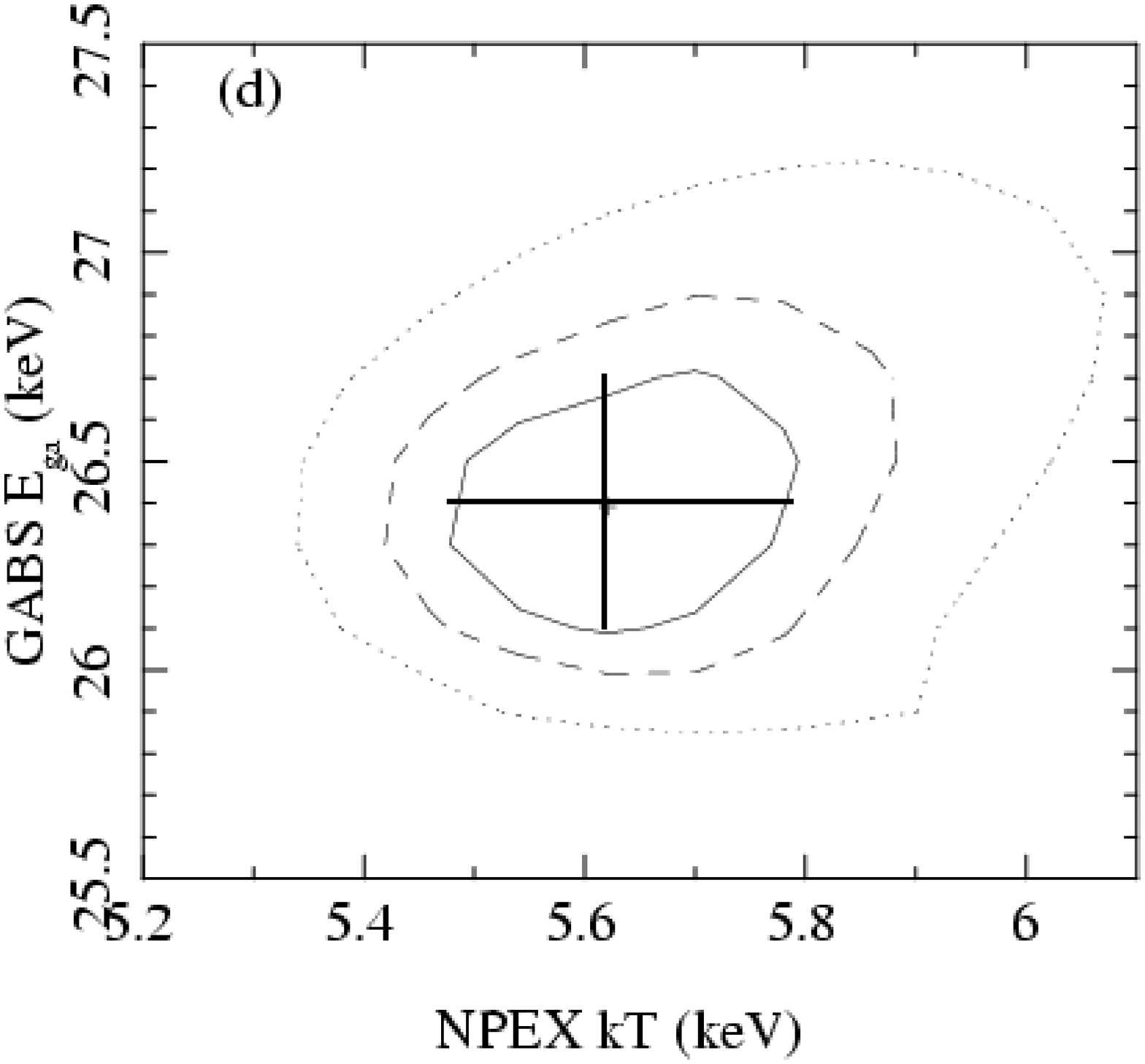}

\plottwo{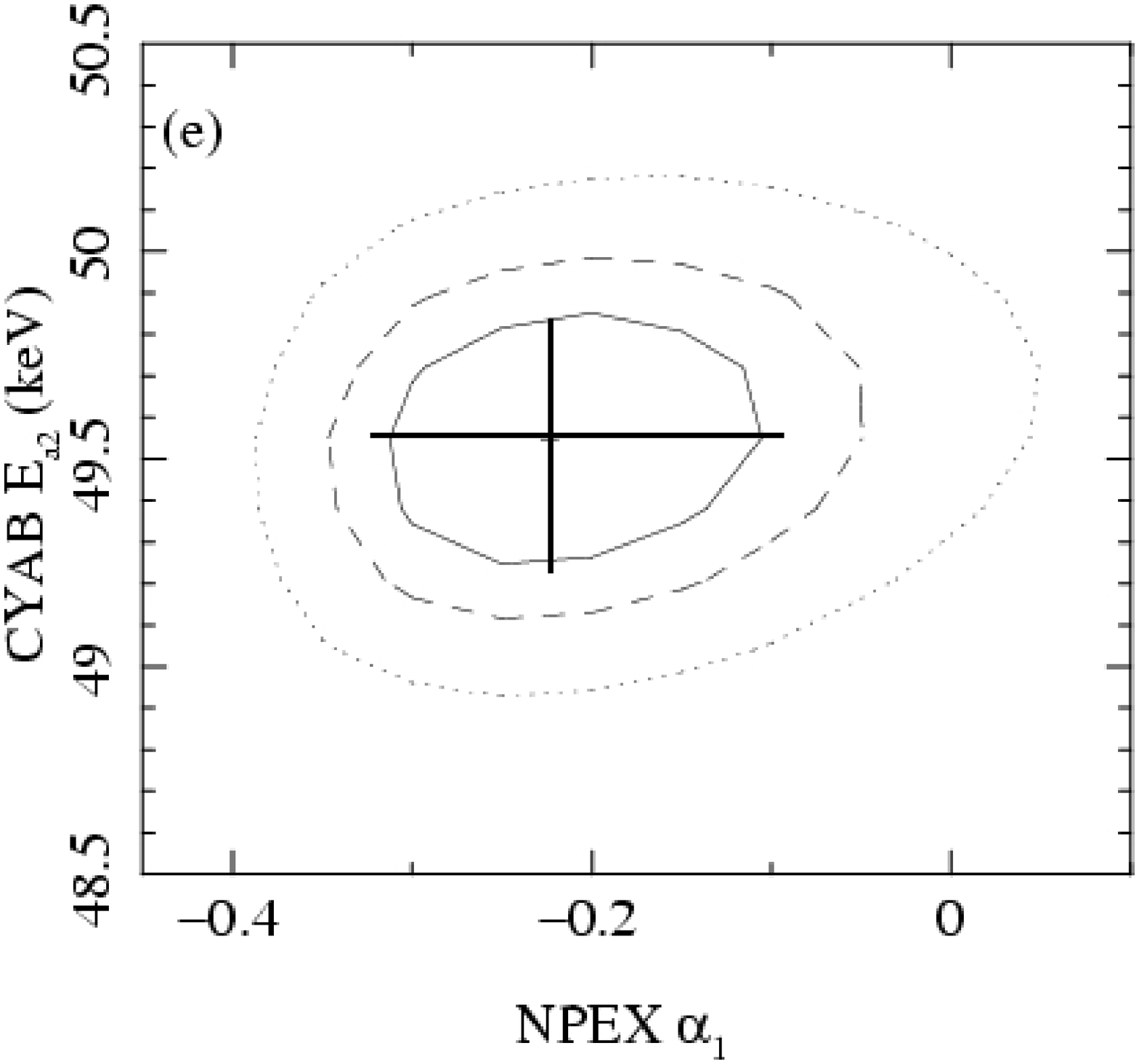}{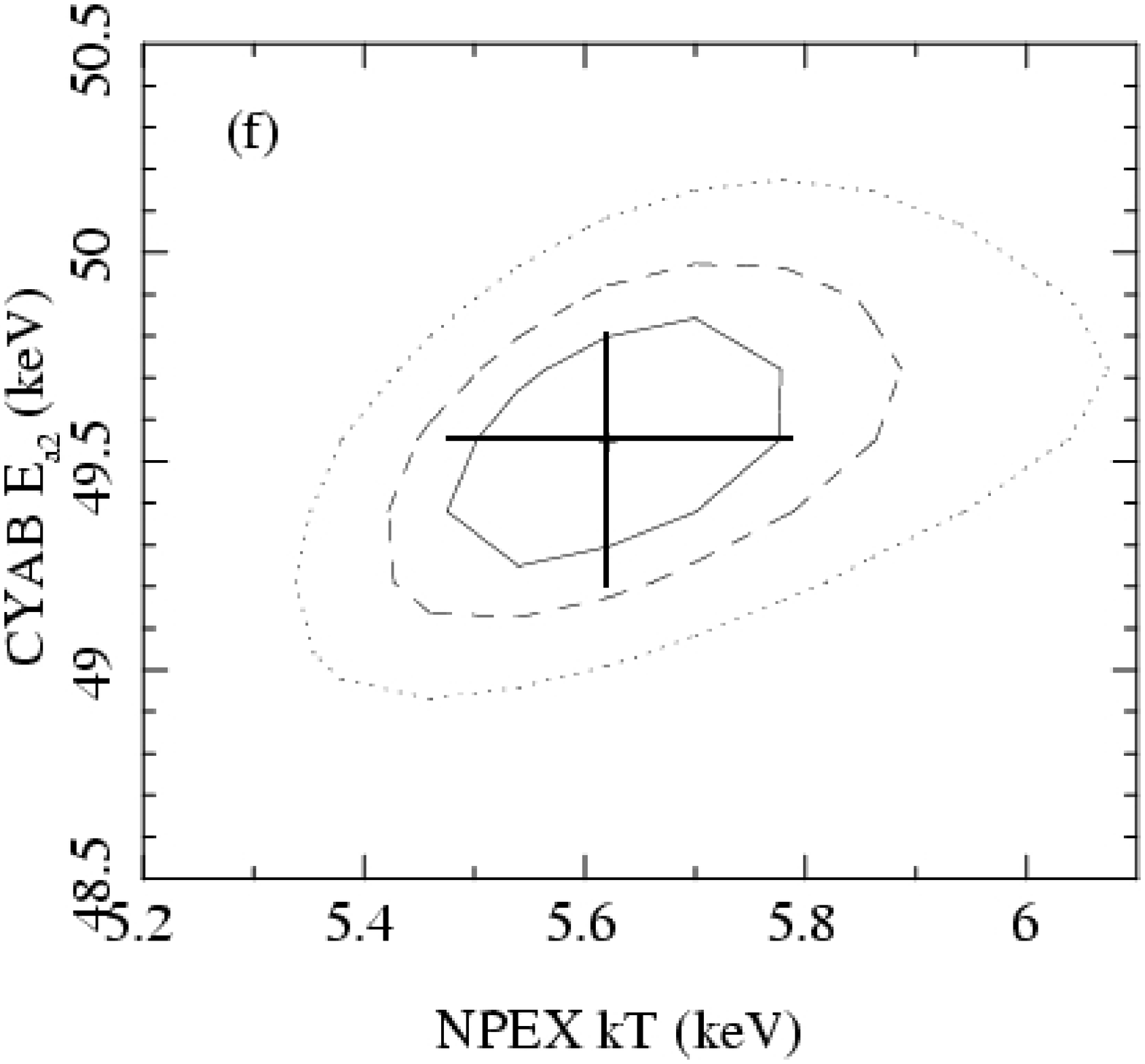}
\caption{Confidence contours between the continuum and the CRSF parameters, obtained from spectral fitting to the Dec 24 data. The solid, dashed, and dotted lines represent 68\%, 90\%, and 99\% confidence levels respectively. The crosses indicate 90\%-confidence error ranges of the individual parameters given in Table 2.}
\label{f4}
\end{figure}

Since the continuum and the CRSF factor couple strongly,
the 90\%-confidence errors of individual fit parameters, given in Table 2,
might be significantly underestimated.
To examine this concern, we present in Figure \ref{f4} 
confidence contours between several pairs of the model parameters,
obtained from the Dec 24 data. 
Thus, $kT$ and the resonance energies are almost uncorrelated.
Although the NPEX $\alpha_1$ and the CYAB {\ea} exhibit some correlations, 
the single-parameter 90\% error ranges are confirmed to 
adequately represent the two-dimensional confidence ranges,
without being under-estimated.

From these spectral analyses, 
we have established the unified model, 
which can reproduce the complex shape of the X0331+53 spectra 
regardless of the source luminosity. 
Hereafter, we use this model, namely, NPEX$\times$CYAB2$\times$GABS.

Although our analysis was performed with the HEAsoft version 6.0.5, 
the PCA response generator has been updated 
in the latest HEAsoft version 6.8, 
and a new PCA CALDB has been released on 2009 December 2. 
To examine possible effects of these software and calibration updates, 
we analyzed the Dec 24 data 
with the HEAsoft version 6.8 and the latest PCA CALDB. 
As a consequence, 
the cyclotron resonance energies changed only $\sim$1.0\%, 
which is within statistical errors. 
Thus, we retain our results obtained with the HEAsoft version 6.0.5.

Just for comparison, 
we attempted to replace the CYAB modeling of the two CRSFs with 
that employing three GABS factors, 
with nested two for the fundamental and the other for the second harmonic. 
The three parameters of the three GABS factors were all left free. 
As a result, the model gave, for example, 
an acceptable fit ($\chi_\nu^2 \sim 1.16$) to the Dec 24 data. 
The obtained CRSF parameters are, 
$E_{1a}=27.2_{-0.4}^{+0.8}$, 
$\sigma_{1a}=8.10_{-1.05}^{+3.39}$, 
$\tau_{1a}=1.46_{-0.28}^{+0.84}$, 
$E_{1b}=25.6_{-0.3}^{+0.2}$, 
$\sigma_{1b}=3.23_{-0.69}^{+0.77}$, 
$\tau_{1b}=0.38_{-0.20}^{+0.49}$, 
$E_{2}=50.2_{-0.5}^{+0.7}$, 
$\sigma_{2}=7.08_{-0.85}^{+0.89}$, and 
$\tau_{2}=1.49_{-0.34}^{+0.34}$. 
Here, the subscript of $1a$ and $1b$ specify 
the parameters of the nested two GABS model 
for the fundamental resonance. 
However, we retain our original modeling using the CYAB model, 
because it gives generally better fits: 
for example, the difference in $\chi^2$ is 28 (for $\nu=51$) 
in the case of the Dec 24 spectrum.

\subsection{Analysis of the Date-sorted Spectra}
\label{sec3.2}

We applied the model established in \S\ref{sec3.1} 
to all the daily-averaged spectra, 
and studied the time evolution of the CRSF parameters. 
Since some data sets have insufficient statistics, 
the GABS model parameters did not converge. 
Following the obtained results in the previous subsection, 
we therefore fixed $E_{{\rm ga}} = 1.2E_{a1}$ in all cases. 
The other procedures of the spectral analyses are 
the same as in \S\ref{sec3.1}.

As a result of this analysis, 
our model has given acceptable fits to 
all the data acquired in the major outburst 
spanning 2004 November through 2005 February. 
While the normalization parameters of the NPEX model 
varied in correlation with the luminosity, 
the NPEX $kT$, which is thought to give a measure of 
the electron temperature in the emission region \citep{max99,tm04}, 
stayed rather constant 
like in the previous results on 4U~0115+63 \citep{m06a,tsy07}. 
Although the fundamental CRSF parameters are well determined, 
the second CRSF parameters in some data sets became unconstrained 
due to insufficient data statistics in higher energies. 
In such cases, we fixed {\eb} at 2{\ea}.

In contrast to the November-February outburst data, 
the remaining three data sets were observed in March 
during the minor outburst recovery (Figure \ref{f1}). 
Since the source was rather faint on these occasions, 
the three spectra were reproduced with 
the NPEX model multiplied by a single CYAB factor 
describing the fundamental CRSF. 
Therefore, we do not discuss these data sets acquired in 2005 March.

\begin{figure}
\epsscale{0.8}
\plotone{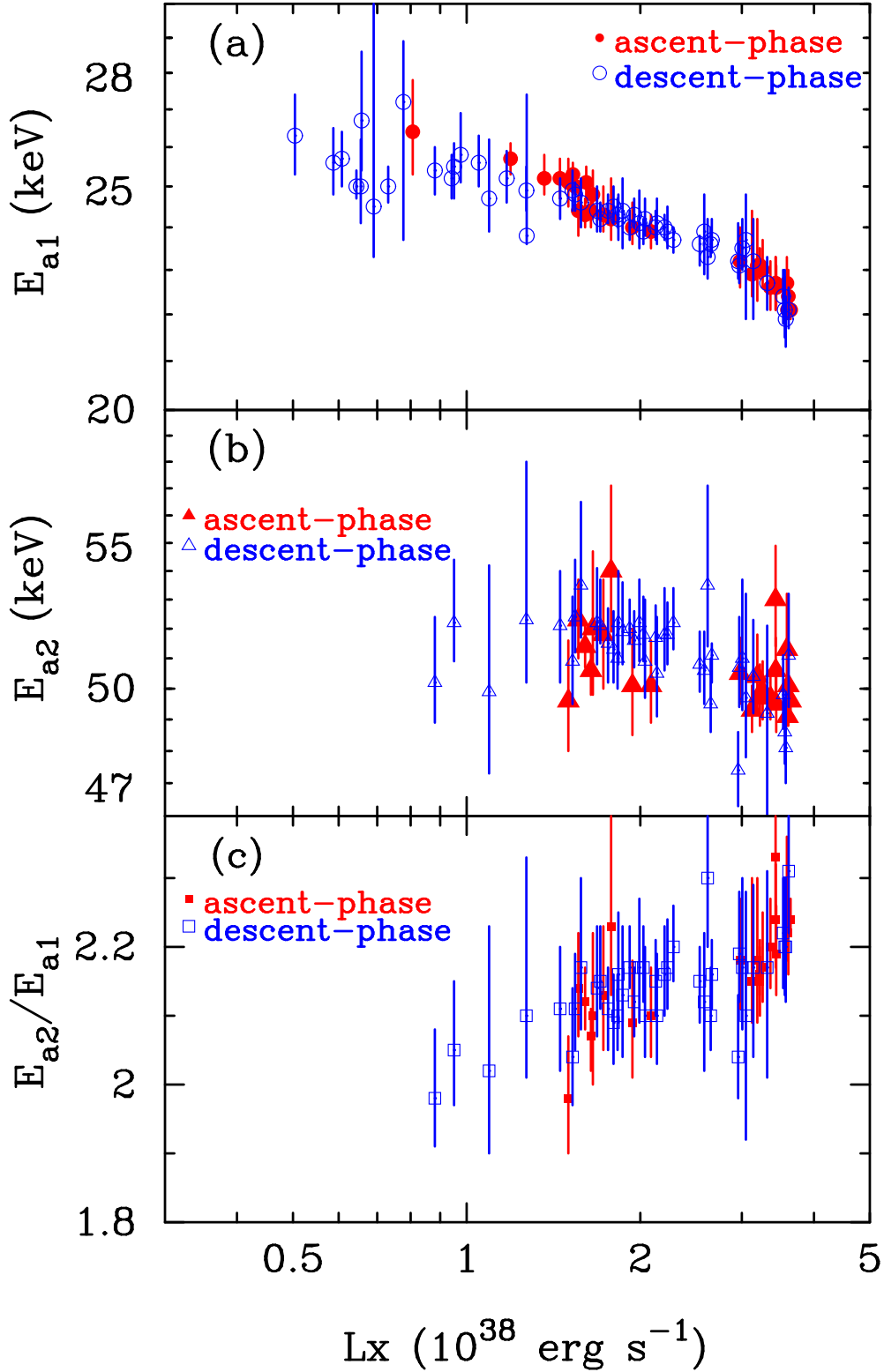}
\caption{The fundamental (panel a) and second harmonic (panel b) cyclotron resonance energies, shown against the $3-80$ keV luminosity. Panel (c) shows the ratios between panels (b) and (a). The plotted data are extracted from the daily-averaged data using the NPEX$\times$CYAB2$\times$GABS model fitting. The error bars represent 90\% confidence levels. Those data in which {\eb} is fixed at 2{\ea} are omitted. The red and blue data represent the ascent and descent portions of the outburst. }
\label{f5}
\end{figure}

Figure \ref{f5}a shows 
the derived values of {\ea}, 
together with 90\%-confidence errors, 
as a function of the $3-80$ keV source luminosity $L_{3,80}$. 
As already reported by Mowlavi et al.\ (2006), Nakajima (2006), 
and Tsygankov et al.\ (2006, 2009), 
{\ea} thus changed from $\sim22$ to $\sim27$ keV 
as $L_{3,80}$ varied between 
$5.0\times10^{37}$ and $3.5\times10^{38}$ {\ergs}. 
When the data points in \ref{f5}a are fitted by a linear function of
$L_{3,80}$
having a form of
\begin{equation}
 E_{\rm a1} = E_{\rm a1}^{(0)} \left\{ 1 + L_{\rm 3,80}/L_1 \right\}~~,
\label{e4}
\end{equation}
its two parameters (with 90\% confidence errors)
were obtained as $E_{\rm a1}^{(0)} =26.5 \pm 0.1$ keV
and $L_1 = (-23.0 \pm 1.2)\times 10^{38}$ {\ergs},
with $\chi_\nu^2=1.3$ ($\nu=81$).
In addition, 
we found no hysteresis effects in the variation of {\ea}
between the ascent and descent phases.
This inference is consistent with a recent report by Tsygankov et al.\ (2009),
who used the same data set. 
Although the values of {\ea} derived here are slightly discrepant with
those measured from the same outburst by other authors
\citep{kre05,pot05,mow06,tsy06},
the difference is within $\sim10$\%, and
can be attributed to the different modelings of the fundamental CRSF
and the continuum.

As shown in Figure \ref{f5}b, 
the second resonance energy {\eb}, when determined independently, 
showed a considerably weaker dependence on $L_{3,80}$. 
When these data points are fitted by another linear function of $L_{3,80}$ as 
\begin{equation} 
 E_{\rm a2} = E_{\rm a2}^{(0)} \left\{ 1 + L_{\rm 3,80}/L_2 \right\}~~, 
\label{e5} 
\end{equation} 
we obtain $E_{\rm a2}^{(0)} =53.5 \pm 0.4$ keV
and $L_2 = (-47.8 \pm 5.6) \times 10^{38}$ {\ergs},
together with $\chi_\nu^2=1.8$ ($\nu=61$).
Thus, $E_{\rm a2}^{(0)}$ is consistent with $2.0E_{\rm a1}^{(0)}$, 
while the luminosity dependence differs significantly between {\ea} and {\eb} 
(i.e., $|L_2| > |L_1|$ beyond their errors).

As a result of the above two correlations,
the {\eb}/{\ea} ratio, shown in Figure \ref{f5}c,
increases from $\sim2.0$ to $\sim2.2$ 
as $L_{\rm 3,80}$ changes from 
$1.0\times10^{38}$ {\ergs} to $3.5\times10^{38}$ {\ergs}. 
Since these ratios give $\chi_\nu^2=2.8$ ($\nu=62$) 
when fitted with a constant value, 
they are inconsistent with being constant. 
In contrast, the fit became satisfactory with $\chi_\nu^2=1.2$ ($\nu=61$), 
when we employ yet another linear function as 
\begin{equation} 
 E_{\rm a2}/E_{\rm a1} = K \left\{ 1 + L_{\rm 3,80}/L_{21} \right\}~~. 
\label{e6} 
\end{equation} 
The two parameters were obtained as 
$K=2.0 \pm 0.1$ and $L_{21} = (34 \pm 5)\times 10^{38}$ {\ergs}.
Thus, the value of $K$ is consistent with 2.0, 
while $L_{21}$ remains finite. 
In other words, the {\eb}/{\ea} ratios exhibit 
a statistically significant dependence on the 3--80 keV luminosity.
In \S3.3 and \S4.2, we discuss possible artifacts on these results 
introduced by our choice of the CRSF modeling.

\subsection{Analysis of the Flux-sorted Spectra }
\label{sec3.3}

\begin{figure}
\epsscale{0.7}
\plotone{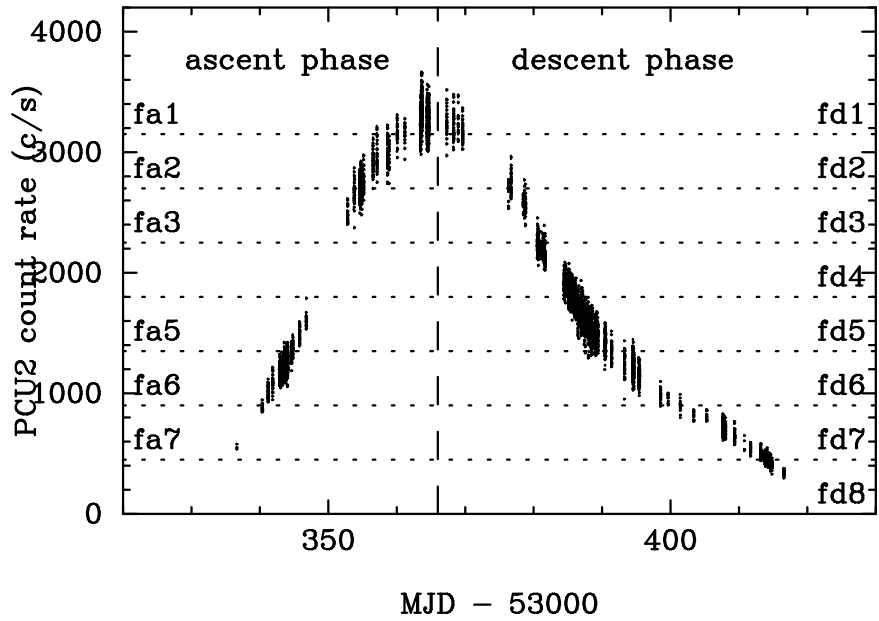}
\caption{The whole PCU 2 lightcurve of X0331+53, plotted with 64 sec binnings. The horizontal dot lines represents boundaries of the flux sorting. The vertical dashed line divides the ascent and descent phases of the outburst. }
\label{f6}
\end{figure}

In order to investigate the behavior of the second CRSF 
under better statistics, 
we next carried out the flux-sorted analysis 
which was already performed successfully on 4U~0115+63 
(Nakajima 2006; Nakajima et al.\ 2006a,b). 
Specifically, 
we sorted the PCU2 and HEXTE cluster B data into 8 flux levels 
in reference to Figure \ref{f6}, 
and co-added those data which fall in the same flux range. 
The flux was calculated every 64 sec, 
so that the flux sorting can catch up with short-term intrinsic variations, 
but not affected by photon-counting statistics. 
In addition, the whole data were divided into 
outburst ascent and descent phases (referring to Figure \ref{f6}), 
to examine the CRSF energy changes for possible hysteresis effects. 
The procedure of the spectral analysis is 
the same as in \S\ref{sec3.2}, 
but $E_{{\rm ga}}$ was left free, because of the improved statistics.

\begin{figure}
\plotone{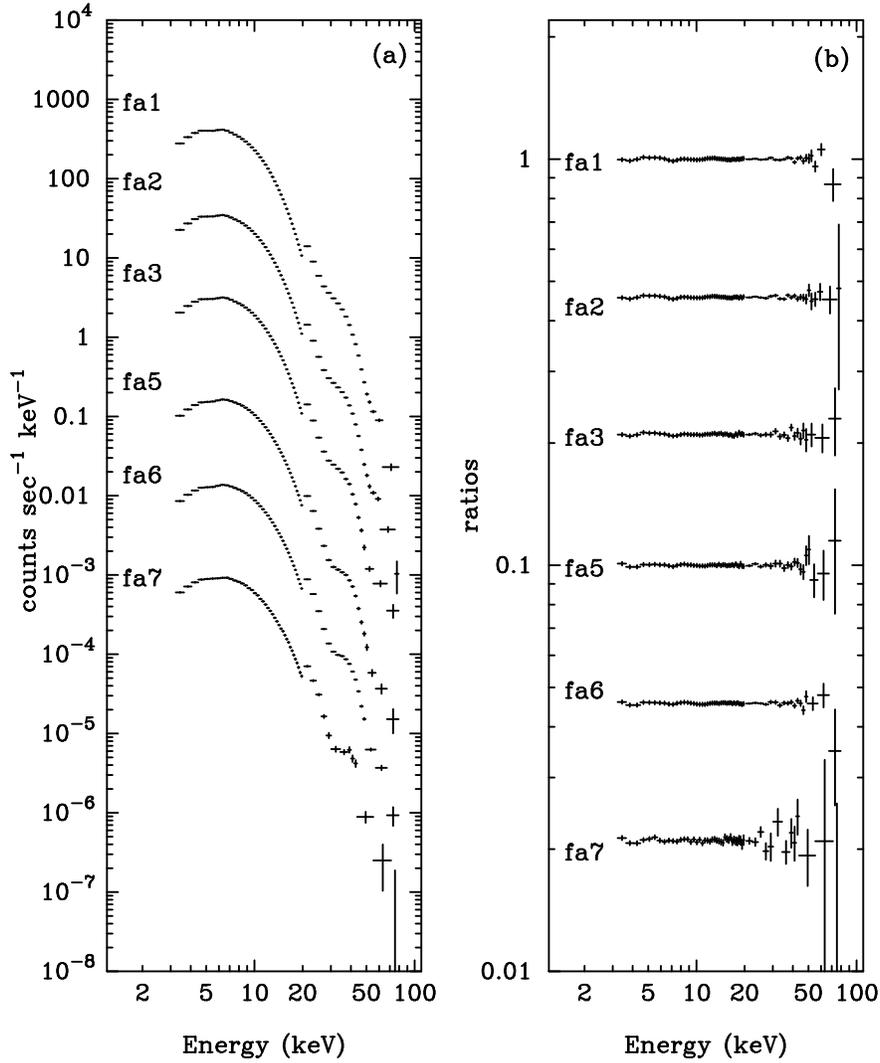}
\caption{(a) Flux-sorted spectra of X0331+53 derived from the ascent phase. For clarify, they are shifted vertically by a factor of 0.5 between neighboring flux levels.  (b) The spectra normalized to the best-fit NPEX$\times$CYAB2$\times$GABS model. The data are shifted vertically, in the same way as in panel (a). }
\label{f7}
\end{figure}

\begin{figure}
\plotone{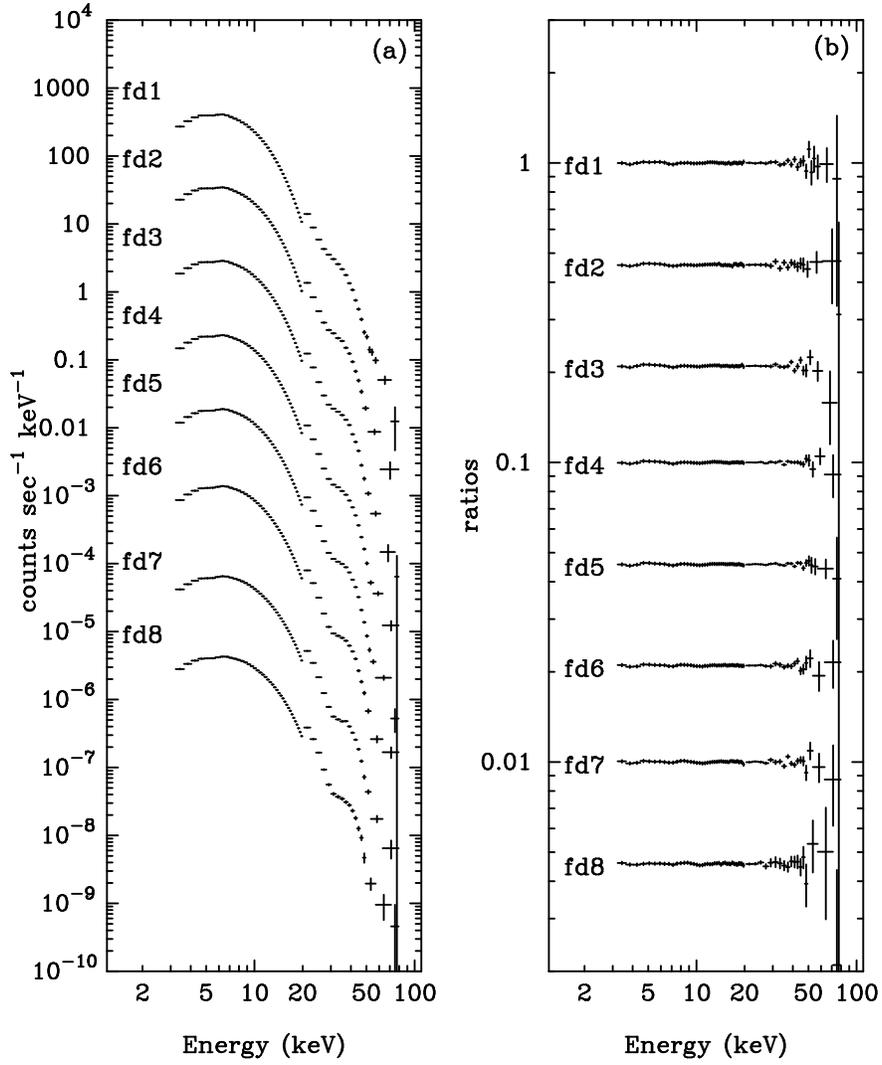}
\caption{(a) Same as Figure \ref{f7}, but the data are derived from the outburst descent phase. }
\label{f8}
\end{figure}

Figure \ref{f7}a and \ref{f8}a show 
the flux-sorted spectra of the ascent- and descent-phases, respectively. 
We again applied the 
NPEX$\times$CYAB2$\times$GABS model to these data, and 
found that the model gives acceptable fits 
as shown in Figure \ref{f7}b and \ref{f8}b.

As summarized in Table \ref{t3},
this analysis allowed us to accurately determine
the parameters of the fundamental and second CRSFs.
Figure \ref{f9}a and b show
the values of {\ea} and {\eb}, respectively, against $L_{\rm 3,80}$.
Thus, {\ea} changed from $\sim22$ to $\sim27$ keV,
while {\eb} from $\sim49$ to $\sim54$ keV, thus
reconfirming the essential features of Figure \ref{f5}
with higher confidence.
In fact, like in the case of the date-sorted spectra (\S3.2),
the fit with eq.\ref{e4} to Figure \ref{f9}a gave
$E_{\rm a1}^{(0)} =27.4 \pm 0.6$ keV,
$L_1 = (-16.5 \pm 3.3)\times 10^{38}$ {\ergs},
and $\chi_\nu^2=0.5$ ($\nu=12$),  
while that with eq.\ref{e5} to Figure \ref{f9}b resulted in
$E_{\rm a2}^{(0)} =53.3 \pm 0.5$ keV,
$L_2 = (-45.7 \pm 7.0)\times 10^{38}$ {\ergs},
and $\chi_\nu^2=1.5$ ($\nu=12$).  
The errors associated with these fit parameters are larger 
than those obtained in the date-sorted analysis,
because here we left $E_{{\rm ga}}$ free
while we previously fixed it at 1.2{\ea}.
Figure \ref{f9}c reconfirms that
the resonance energy ratio {\eb}/{\ea} changed from $\sim2.2$ to
the nominal value $\sim2.0$ as the source luminosity decreased.
The fit with eq.\ref{e6} indicates
$K =2.0 \pm 0.1$,
$L_{21} = (21.2 \pm 7.0)\times 10^{38}$ {\ergs},
and $\chi_\nu^2=0.3$ ($\nu=12$).

\begin{figure}
\plotone{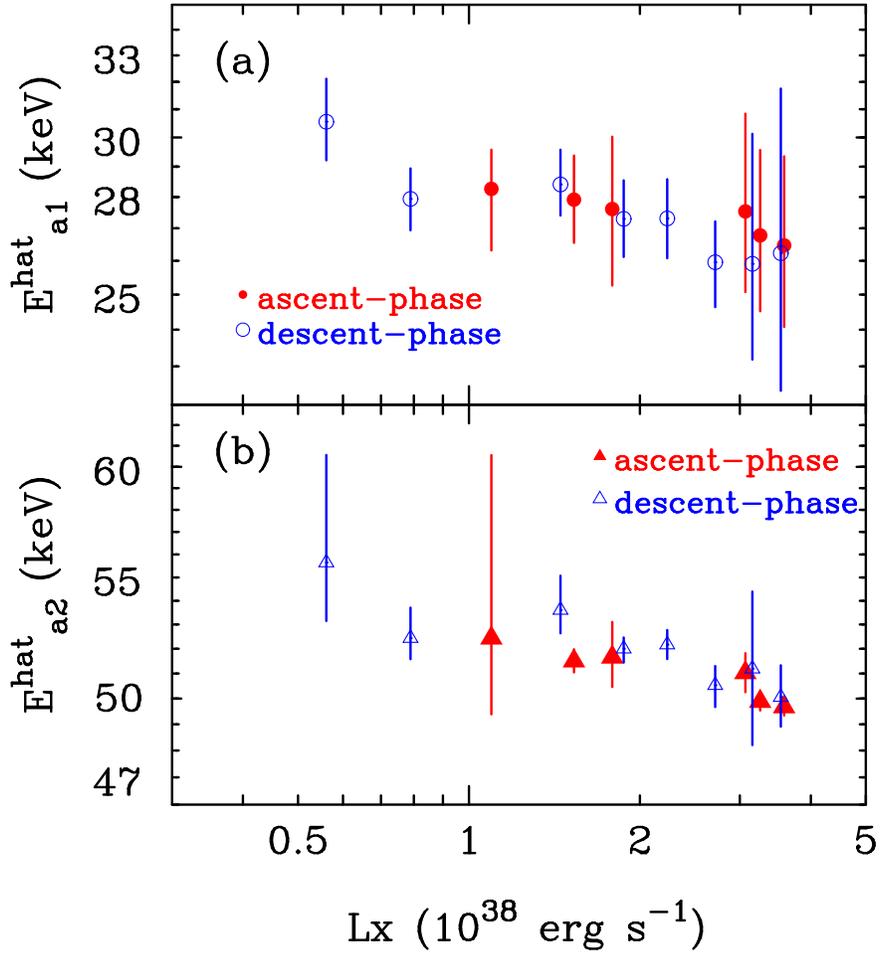}
\caption{The luminosity dependence of $\hat{E}_{{\rm a}1}$ and $\hat{E}_{{\rm a}2}$. }
\label{f10}
\end{figure}

Through the date-sorted and flux-sorted analyses, 
we confirmed that the luminosity-dependent changes in  
{\eb}, {\ea} and the {\eb}/{\ea} ratio are statistically significant.
Then, does this conclusion remain unaffected 
even if considering systematic errors?
An immediate concern, namely spurious effects 
due to coupling between the continuum and CRSF parameters, 
can be ruled out because 
the error estimates are appropriate as indicated by Figure~\ref{f4}.
Another problem is that the values of {\ea} and {\eb}
do not pick up the deepest position of the CRSFs.
Accordingly, we examined the behavior of the energy 
$\hat{E}_{{\rm a} i} \equiv E_{{\rm a} i} \left\{ 1 + (W_i/E_{{\rm a} i})^2 \right\}$ ($i=1,2$) 
at which the absorption is expected to become deepest \citep{tm95}.
Figure~\ref{f10} shows luminosity dependence of $\hat{E}_{{\rm a} i}$. 
Thus, the values of $\hat{E}_{{\rm a} i}$ also 
exhibit a clear luminosity dependence. 
By fitting these data points with eq.\ref{e4} and eq.\ref{e5}, 
we obtained the characteristic luminosities as 
$\hat{L_1} = (-27.0 \pm 11.8)\times 10^{38}$ {\ergs} and 
$\hat{L_2} = (-44.3 \pm 8.0) \times 10^{38}$ {\ergs}.
We hence reconfirm that the second resonance energy depends 
less steeply on the luminosity than the fundamental energy, 
although the difference between $\hat{L}_1$ and $\hat{L}_2$ is less
significant than that between $L_1$ and $L_2$, 
and the average $\hat{E}_{\rm a2}/\hat{E}_{\rm a1}$ ratio, 
1.8--2.0 is smaller than 2.0
in contrast to the case of the {\eb}/{\ea} ratios.

\begin{figure}
\plotone{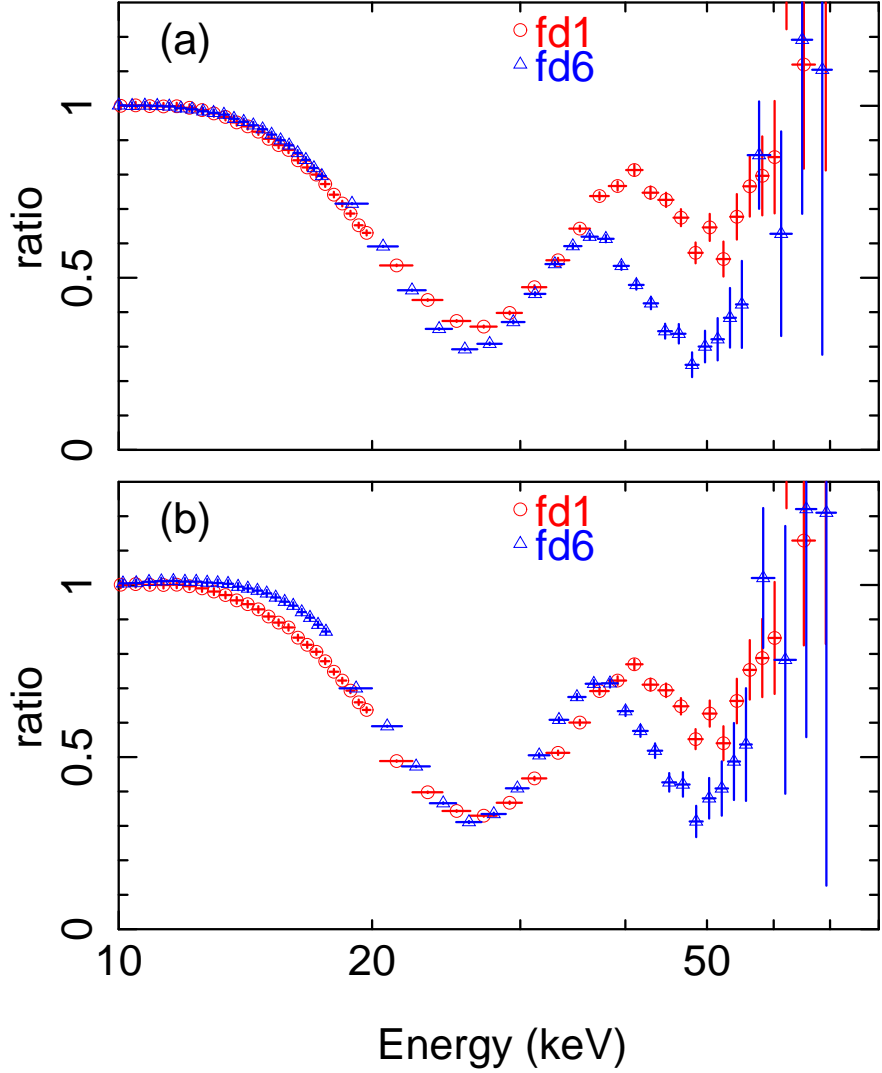}
\caption{(a) The flux-sorted fd1 (red circles) and fd6 (blue triangles) spectra, normalized to the best-fit NPEX continuum determined individually using the 3--13 and 60--80 keV ranges only. The energy scale of the blue spectrum is scaled by a factor of 0.885, and it is compressed in the vertical direction by a factor of 1+(1-y)$\times$0.38 (with the ratio 1.0 kept as a pivot). The $<20$ and $>20)$ keV ranges are covered by the PCA and HEXTE data, respectively.  (b) The same as panel (a), but the PLCUT model is used as the continuum instead of NPEX. Again, the two continua are normalized to separate models. With respect to the red data points, the blue ones are scaled by a factor of 0.895 in the energy axis, and 1+(1-y)$\times$0.40 in the vertical direction. }
\label{f11}
\end{figure}

The largest systematic effect that may possibly affect the {\eb}/{\ea}
ratio would be the choice of the continuum and CRSF models.
Since we have already tried in \S\ref{sec3.1}
replacing the CRSF model with the GABS model,
here we examine the choice of the continuum.
In order not to be affected by the CRSF modeling,
we fitted the lowest-energy (3-13 keV) 
and highest-energy (60--80 keV) ends of the flux-sorted fd1 and fd6 spectra,
using a single NPEX continuum,
and then normalized the overall spectrum to the best-fit model
that was determined by each spectrum.
The results, shown in Figure \ref{f11}a, 
reveal the absorption features in a relatively model-free manner,
with the continuum approximately removed.
There, we have adjusted the energy scale and the ratio scale
between the two spectra,
so that the fundamental CRSFs best overlap.
Thus, the fd1 ratio spectrum exhibits the 2nd harmonic 
at a significantly higher energy than the fd6 ratio.
This result visualize the luminosity-dependent change in the {\eb}/{\ea} ratio.

To examine the overall behavior of {\eb}, {\ea},
and their ratios under a different continuum model,
we employed a power-law cutoff (PLCUT) model (Coburn et al.\ 2002),
namely a power-law multiplied by an exponential cutoff factor.
Figure~\ref{f11}b represents the same analysis as Figure~\ref{f11}a
conducted with this continuum model.
Thus, the result again visualize
that fd1 has a higher {\eb}/{\ea} ratio than fd6.
We further fitted the fd1 and fd6 spectra
by multiplying these PLCUT continua
with the same absorption factors as used so far,
and obtained the results summarized in Table \ref{t4}.
Thus, the fit goodness is comparable to
those obtained with the NPEX continuum, and the values
of {\ea} and {\eb} remain unchanged within errors,
although the luminosity-dependent change in the
{\eb}/{\ea} ratio became less significant.

As our further confirmation of the luminosity-dependent
changes in the CRSF energies and their ratios,
we fitted the fd1 and fd6 spectra simultaneously,
with the NPEX$\times$CYAB2 model.
While the model parameters were generally
allowed to differ between the two spectra,
we imposed 4 stepwise constraints on the resonance energies.
In the first step, we required {\ea} to be the same
between fd1 and fd6, and constrained {\eb} to be twice {\ea}.
However, the fit was not acceptable, with $\chi^2/\nu=2365/111$.
Next, we allowed fd1 and fd6 to have different values of {\ea},
but retained the 1:2 harmonic constraint on both spectra.
This gave an improved fit with $\chi^2/\nu=398/110$,
implying that the two spectra have different resonance energies.
Third, we allowed the two spectra to have different values of {\ea}, 
but required them to have the same {\eb}/{\ea} ratio.
The fit was then improved to $\chi^2/\nu=321/110$,
together with {\eb}/{\ea}$ = 2.11\pm0.08$
which deviate from the the nominal value 2.0. 
Finally, we left all the four resonance energies free,
and obtained a improved fit with $\chi^2/\nu=282/108$. 
This means that the {\eb}/{\ea} ratio is different not only 
from the harmonic condition, but also between the two spectra.
(The fit is still unacceptable because we did not include 
the GABS factor to model the narrow core of the 
fundamental resonance.)

Finally, we repeated the same 4-step analyses,
but using the PLCUT continuum.
Then, the fit goodness from the 4 steps were
$\chi^2/\nu=2091/111$, $\chi^2/\nu=1119/110$, $\chi^2/\nu=341/110$,
and $\chi^2/\nu=254/108$. 
The implication is essentially the same as that
derived with the NPEX continuum.

\subsection{Additional Analyses}
\label{sec3.4}

Since the flux-sorted spectra have rather good statistics, 
we examined them, up to 110 keV, 
for evidence of the third harmonic CRSF 
which was already reported by 
Kreykenbohm et al.\ (2005) and Pottschmidt et al.\ (2005). 
For this purpose, 
we further multiplied the NPEX$\times$CYAB2$\times$GABS model 
by an additional CYAB factor centered at $\sim75$ keV, 
and applied it to the flux sorted spectra. 
Then, two data sets, fa1 and fd2, 
gave statistical significant (at $\sim$90\%) 
evidence for the third CRSF, 
while the fundamental and second-harmonic CRSF parameters 
remained essentially the same as before. 
As a representative case, 
the third resonance parameters derived from the fa1 data are 
{\ec}$=73^{+3}_{-2}$ keV, 
$W_3=6^{+7}_{-5}$ keV, 
and $D_3=3^{+23}_{-1}$. 
These  parameters are comparable to 
the previous results derived from the same {\xte} data \citep{kre05,pot05}. 
The third resonance energy was found 
to change no less than by $\sim \pm$3\%, 
as the $3-80$ keV luminosity varied from 
$3.16$ to $3.59\times10^{38}$ {\ergs}. 
Over this luminosity range, 
the {\ec}/{\eb} ratio 
stayed at an average value of 1.46 
within a typical uncertainty of $\pm 0.05$. 
In contrast, the {\ec}/{\ea} ratio is systematically higher, 3.2. 
Therefore, {\ec} and {\eb} are 
consistent with being in the 2:3 harmonic ratio, 
while {\ea} is inferred to be systematically lower, 
and depend significantly on the luminosity.

In addition to the investigation of the resonance energy changes,
we examined whether the depths of the cyclotron resonance,
$D_{i}$ ($i=1,2$) and $\tau_{{\rm ga}}$, depend on the luminosity.
Figure \ref{f12} shows these depth parameters,
against the $3-80$ keV luminosity.
Both the fundamental and second CRSF depths, $D_1$ and $D_2$
respectively
were found to decrease toward higher luminosities.
Although $\tau_{{\rm ga}}$ has large errors,
it also exhibits the same tendency.
These luminosity dependence in $D_1$ and $D_2$ are statistically significant, 
because fitting the $D_1$ and $D_2$ measurements with constant
values gave reduced chi-square of 
2.11 ($\nu=13$), 5.88 ($\nu=13$), respectively.

\section{DISCUSSION}
\label{sec4}

\subsection{Modeling of the Spectra}
\label{sec4.1}

We analyzed the whole set of {\xte} PCA and HEXTE data of X0331+53, 
covering the 2004$-$2005 outburst, with two objectives in mind. 
One is to examine 
whether the luminosity-anticorrelated changes 
in the fundamental CRSF energy, 
observed during the outburst descent phase 
(Mowlavi et al.\ 2006; Nakajima 2006; Tsygankov et al.\ 2006), 
is also present during the outburst rising phase. 
This agrees with the conclusion of Tsygankov et al.\ (2009). 
The other is to examine the reported second harmonic resonance of this pulsar 
(Coburn et al.\ 2005; Kreykenbohm et al.\ 2005; Pottschmidt et al. \ 2005) 
for possible luminosity-dependent changes, 
and compare the results with those of the fundamental CRSF.

Following our experience 
(e.g., Mihara 1995; Makishima et al.\ 1999), 
we adopted the NPEX model of eq.\ref{e1} to represent 
the underlying 3$-$80 keV PCA and HEXTE continua, 
while two CYAB factors, eq.\ref{e2}, to account for 
the fundamental and second harmonic cyclotron features. 
Furthermore, to express the complex shape 
of the deep fundamental cyclotron feature 
as revealed in the previous studies \citep{max90a,kre05,pot05}, 
we introduced a fine-tuning GABS factor, eq.\ref{e3}, 
which is nested with the fundamental CYAB factor. 
This NPEX$\times$CYAB2$\times$GABS model has given 
acceptable fits to all the daily-averaged and flux-sorted spectra. 
Furthermore, some of the spectra exhibited 
statistically significant evidence for the third harmonic resonance, 
as already reported by Coburn et al.\ (2005), Kreykenbohm et al.\ (2005), 
and Pottschmidt et al.\ (2005).

As exemplified in Figure \ref{f12} and Table \ref{t3}, 
our fits generally imply 
that the fundamental and second harmonic resonances 
both have optical depths of $1.0-2.0$, 
with comparable CYAB widths of $\sim 8$ keV. 
The fundamental CRSF is explained mainly by the CYAB model, 
with the additional GABS factor 
carrying $\sim 30\%$ of the optical depth 
when the source is luminous. 
Replacing the two CYAB models with two GABS models 
made the fits significantly worse 
(e.g. {\ch}$=3.6$ in fa2 spectrum fitting), 
mainly because the observed flux falls 
considerably below the NPEX continuum, 
even at the $35-45$ keV range 
which is in between the two cyclotron troughs 
(see Figure \ref{f3}). 
This effect cannot be reproduced adequately by the 
GABS model which is symmetric between the red and blue sides, 
while it can be successfully accounted for by the CYAB factor 
of which the wing is more extended in the blue side 
due to the $E^2$ factor in eq.\ref{e2}.

\subsection{Changes of the Cyclotron Resonance Energies}
\label{sec4.2}
 
By applying the NPEX$\times$CYAB2$\times$GABS model 
to the daily-averaged and flux-sorted PCA$+$HEXTE spectra, 
we reconfirmed the reports by Mowlavi et al.\ (2006), 
Nakajima (2006), and Tsygankov et al.\ (2006, 2009), 
that the fundamental CRSF energy {\ea} decreases 
as the source gets more luminous (Figure \ref{f9}). 
Thus, X0331+53 becomes a second binary X-ray pulsar, 
after 4U~0115+63, of which {\ea} correlates 
negatively with the source luminosity. 
Like the case of 4U~0115+63, 
this effect observed from X0331+53 can be explained by presuming 
that the cyclotron resonance ``photosphere'' 
gets higher in the accretion column, 
as the source luminosity increases 
and hence the column becomes taller \citep{bas76,bur91,tm04}. 
These cases make a contrast to the behavior of Her X-1, 
in which {\ea} is reported to have been varying 
in a positive correlation with the luminosity \citep{gru01,sta07}, 
at least while the source is in a sub-Eddington regime. 
On the other hand, 
the recent works of Her X-1 \citep{sta07,klo08} reported 
that the anti-correlation behavior might exist
when the source was in the super-Eddington regime.

In 4U~0115+63, 
the luminosity dependence of {\ea} 
showed some hysteresis effects 
between the ascent and descent phases of the outburst 
(Nakajima et al.\ 2006). 
In contrast, 
we found, in agreement with Tsygankov et al.\ (2009), 
no such effects from X0331+53; 
in this source, 
the value of {\ea} can be considered 
as a single-valued function of the source luminosity.

As the most important result of the present work, 
our NPEX$\times$CYAB2$\times$GABS modeling leads to an
inference that
the second resonance energy {\eb} also 
correlates negatively with the source luminosity, 
but significantly more weakly than {\ea} does. 
The two resonance energies decrease by 
$\Delta E_{\rm a1}/E_{\rm a1}\sim12\%$ 
and $\Delta E_{\rm a2}/E_{\rm a2}\sim5\%$, 
as the $3-80$ keV luminosity increases from 
$1.0\times10^{38}$ to $3.5\times10^{38}$ {\ergs}. 
Consequently,
the {\eb}/{\ea} ratio 
is consistent with the nominal value of 2.0 
when the source is dim, 
while it increases to $\sim2.2$ 
toward the outburst peak (Figure \ref{f9}c). 
These effects are
somewhat model dependent, and become less significant
when the absorption profile and/or continuum shape
are modeled in different ways.
Although these effects were already noticed by Tsygankov et al.\ (2006),
their spectra had rather low statistics in higher energies,
and hence the second CRSF parameters were not well constrained.

While the relativistic effects \citep{ara99} predicts 
the {\eb}/{\ea} ratios to be lower than the harmonic value of 2, 
actual measurements do not necessarily agree.
The ratio exceeds 2.0 in some cases,
and fall in the range $2.1\sim2.2$;
for example,  4U~1907+09 showed {\ea} $\sim18$ keV 
and {\eb} $\sim38$ keV \citep{cus98}; 
Vela X-1 showed {\ea} $\sim24$ keV 
and {\eb} $\sim52$ keV \citep{max99,kre02}; 
and A0535+26 {\ea} $\sim50$ keV 
and {\eb} $\sim110$ keV \citep{kre06}. 
In contrast, the ratios lower than 2.0 are reported in other cases,
including the present {\it RXTE} data of X0331+53 themselves
analyzed by some other authors (e.g., Pottschmidt et al.\ 2006) 
who generally took the deepest positions of the CRSFs 
as the resonance energies (\S\ref{sec3.3}).

A dominant origin of the reported scatter in the {\eb}/{\ea} ratio 
is presumably
the use of different (e.g., CYAB vs. GABS) modeling of the CRSF profile. 
In this sense, 
our result of {\eb}/{\ea}$>2$ is not free from this systematic problem. 
In fact, we find {\eb}/{\ea}$<$2 
if instead using $\hat{E}_{{\rm a} 2}$/$\hat{E}_{{\rm a} 1}$. 
Nevertheless, our result has some pieces of supporting evidence.
One is that 
the CYAB modeling is more successful than that with the GABS
factor (\S\ref{sec3.1}), when combined with the NPEX continuum.
Another is that the third to second resonance energy ratio, 
{\ec}/{\eb}, is consistent with 1.5, 
while {\ec}/{\ea} is higher than the nominal value of 3.0. 
Yet another support is provided by a physically self-consistent
interpretation to be described in the next subsection.
Even admitting that the absolute values of {\ea} and {\eb} 
might be subject to the systematic modeling uncertainly, 
the luminosity-dependent changes in the resonance energy ratio 
is considered to be robust as already shown in \S\ref{sec3.3}.

\subsection{A Possible Explanation of the Observed Effect}
\label{sec4.3}

Let us consider a possible explanation of the 
luminosity-dependent changes in {\ea}, {\eb}, and their ratios. 
According to Mihara et al.\ (2004) and Nakajima et al.\ (2006a), 
the observed changes in {\ea} of 4U~0115+63 
can be explained in terms of those in 
the height of cyclotron resonance ``photosphere'', 
in combination with a dipole field geometry. 
In this scenario, the photosphere height 
above the neutron star surface, $h_r$, 
can be estimated as 
\begin{equation} 
  \frac{h_{\rm r}}{R_{\rm NS}} \approx \ 
  \left( \frac{E_{{\rm a} i}}{E_{{\rm s} i}} \right)^{-1/3} - 1 ~, 
\label{e7} 
\end{equation} 
where $R_{\rm NS}$ is the neutron star radius, 
$E_{{\rm s} i}$ is the resonance energy 
to be observed on the neutron-star surface, 
and $i$ is the harmonic number.

\begin{figure}
\plotone{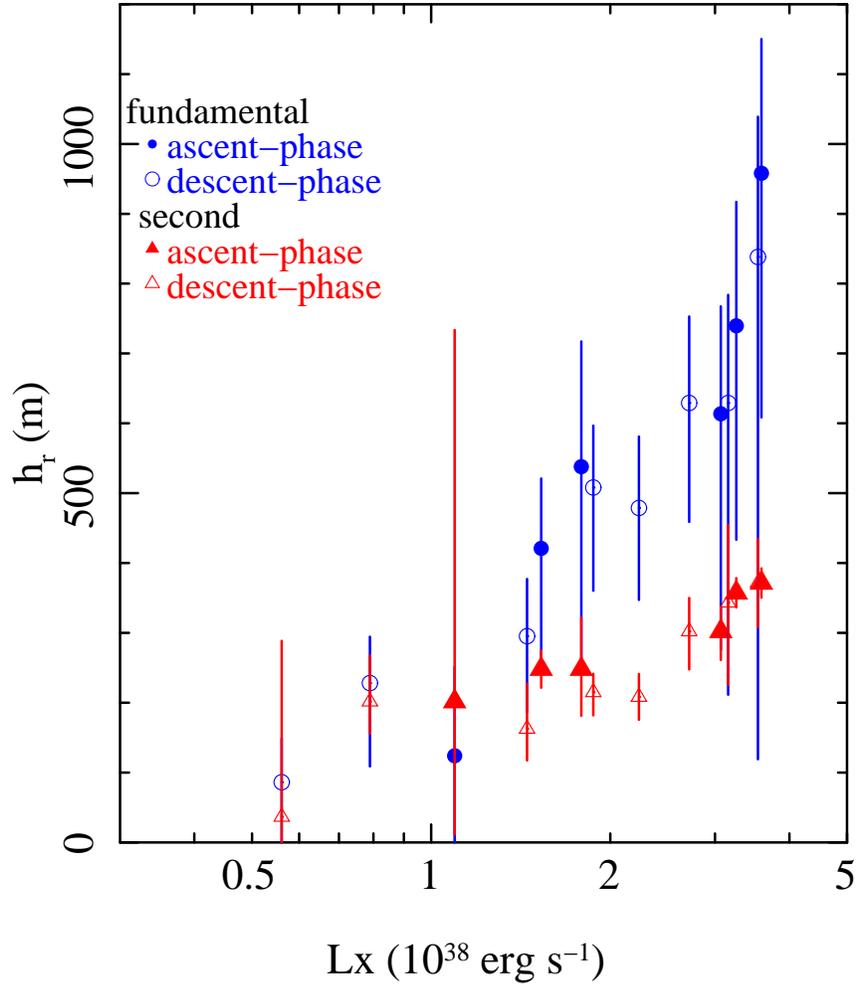}
\caption{The estimated height of the CRSF forming regions against the $3-80$ keV luminosity. }
\label{f13}
\end{figure}

Substituting the {\ea} and {\eb} values derived from 
the flux-sorted analysis (\S\ref{sec3.3}) into eq.\ref{e7}, 
$h_r$ of the fundamental and second resonances 
have been estimated as shown in Figure \ref{f13}. 
Here, we employed $E_{{\rm s} 1} = 27$ keV and $E_{{\rm s} 2} = 54$ keV, 
referring to the lowest-luminosity state in Figure \ref{f9}. 
Also a typical value of $R_{\rm NS} = 10$ km was employed. 
By translating the measurements into the photospheric heights, 
the figure thus yields two important implications. 
One is that the photosphere of the fundamental CRSF 
gets higher as the source luminosity increases, 
up to 800 m, or $\sim 8$\% of $R_{\rm NS}$. 
The other is that the second resonance photosphere, 
though increasing with the luminosity as well, 
is located closer to the surface than that of the fundamental, 
just reflecting the deviation of the {\eb}/{\ea} ratio from 2.0. 
The height difference between the two photospheres reach $\sim$500 m 
when the source is most luminous. 
Like the estimation of $h_r$ described above,
we also calculated the resonance heights using the asymptotic
energies $\hat{E}_{{\rm s} i}$ which are derived from $E_{{\rm a} i}$.
Assume $\hat{E}_{{\rm s} 1}=31.0$ keV and 
$\hat{E}_{{\rm s} 2}=62.0$ keV,
the luminosity-dependent height variations turned out to be
$\sim500$ m for $\hat{E}_{{\rm a} 1}$ and
$\sim400$ m for $\hat{E}_{{\rm a} 2}$.

The altitude difference between the two photospheres 
can be explained qualitatively in the following manner. 
Theoretically \citep[and references therein]{har91,ara99}, 
the fundamental CRSF is predicted to have 
a cross section which is $\sim10$ times larger 
than that of the second resonance. 
Then, we expect the fundamental photosphere to be formed 
nearly at the top of the accretion column. 
In contrast, 
the second harmonic photosphere will appear at lower altitudes, 
where the column density integrated along 
our line of sight is expected to become higher. 
These effects have already predicted theoretically by Nishimura (2008). 
Then, a detailed modeling of the {\eb}/{\ea} ratio 
may provide a valuable probe into the density distribution 
along the accretion column, 
and its dependence on the mass accretion rate, 
although such a work is beyond the scope 
of the present paper.

Assuming that the luminosity-dependent changes in the 
resonance energies are caused by the photospheric
height variations, and not, e.g., by relativistic
effects (\S\ref{sec4.2}), 
the above physical picture further gives an {\it a posteriori} support
to our finding of {\eb}/{\ea}$>2.0$.
In fact, if this ratio fell below 2.0 toward lower luminosities,
we would have to conclude
that the 2nd resonance photosphere is
located higher than that of the fundamental,
leading to a physically unrealistic condition.

\subsection{Comparison with 4U~0115+63}
\label{sec4.4}

\begin{figure}
\plotone{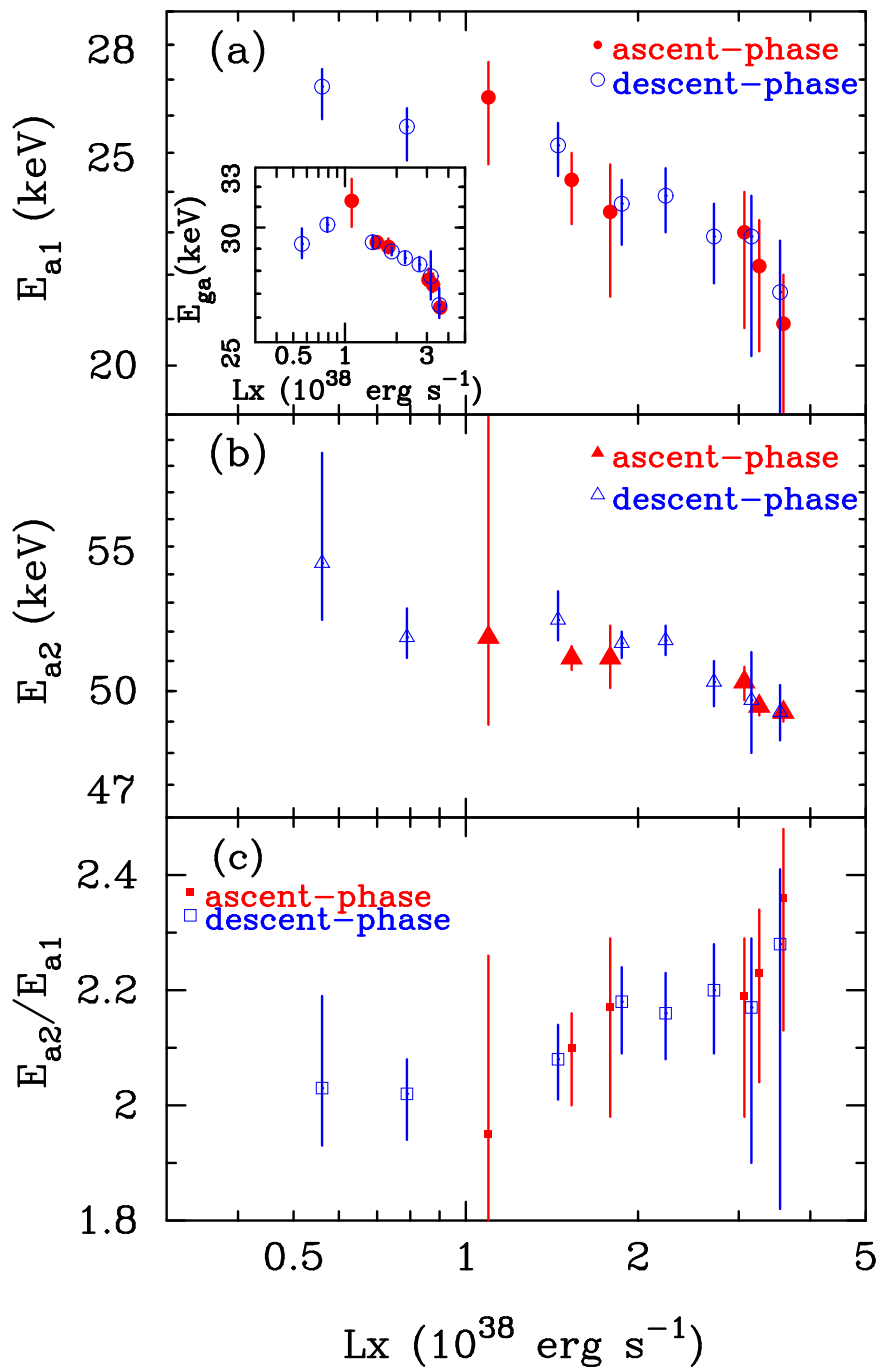}
\caption{Same as Figure \ref{f5}, but the results are derived using the flux-sorted spectra.  The inset panel in (a) describes the luminosity dependence of the $E_{{\rm ga}}$ parameter. }
\label{f9}
\end{figure}

\begin{figure}
\plotone{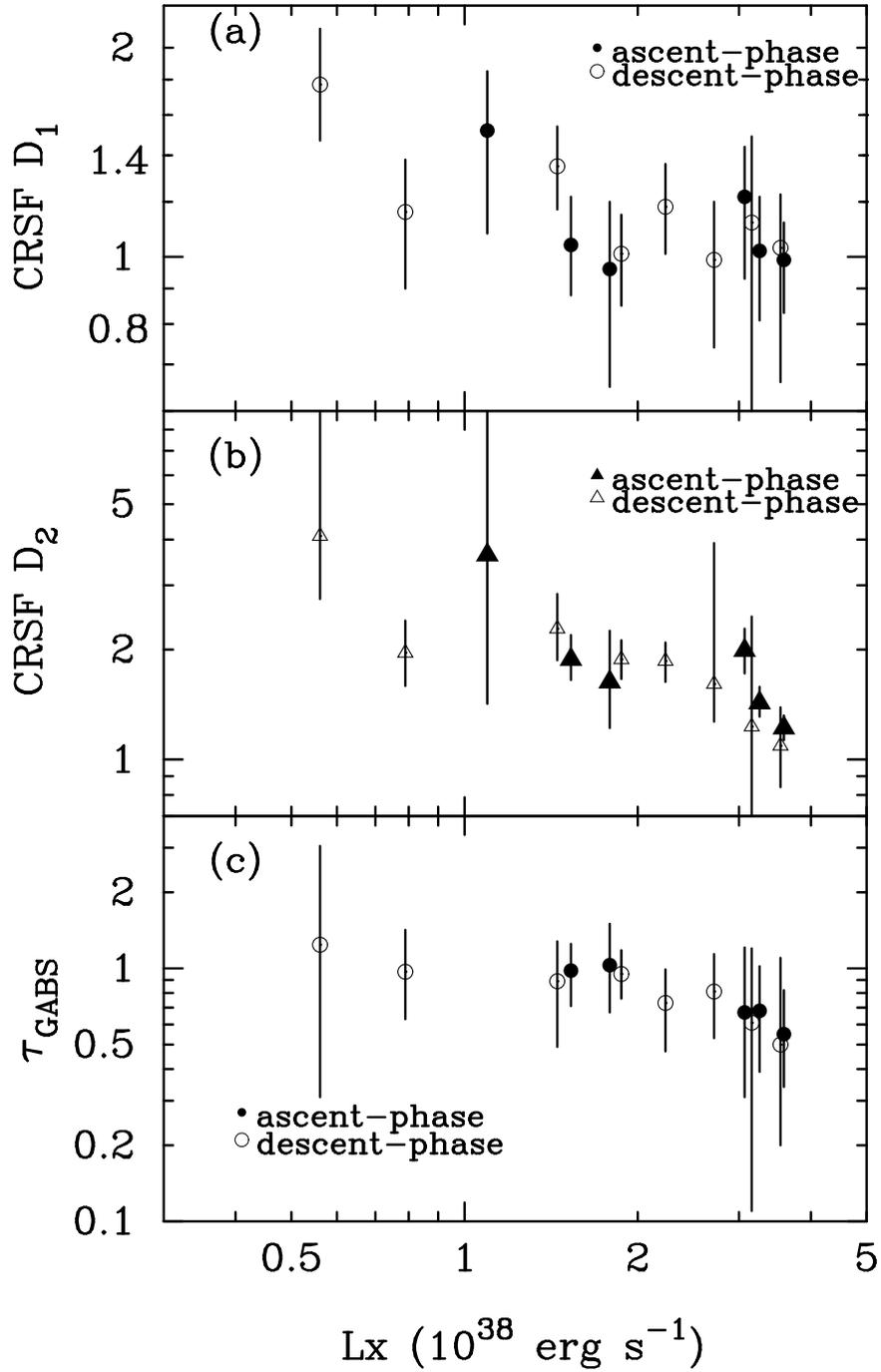}
\caption{The CRSF depths derived from the flux-sorted spectra, plotted against the $3-80$ keV source luminosity. }
\label{f12}
\end{figure}

Following our previous results on 4U~0115+63 \citep{tm04,m06,m06a}, 
the present work provides valuable information 
as to the luminosity dependent changes of the CRSF parameters 
in accreting X-ray pulsars. 
Although the fundamental CRSF of X0331+53 required 
the additional GABS factor (which was not needed in 4U~0115+63), 
the CYAB and GABS centroids varied similarly 
as shown in Figure \ref{f9} and \ref{f12}. 
We therefore consider 
that this slight difference in the CRSF modeling 
does not hamper a direct comparison 
of the results from the two sources.

\begin{figure}
\plotone{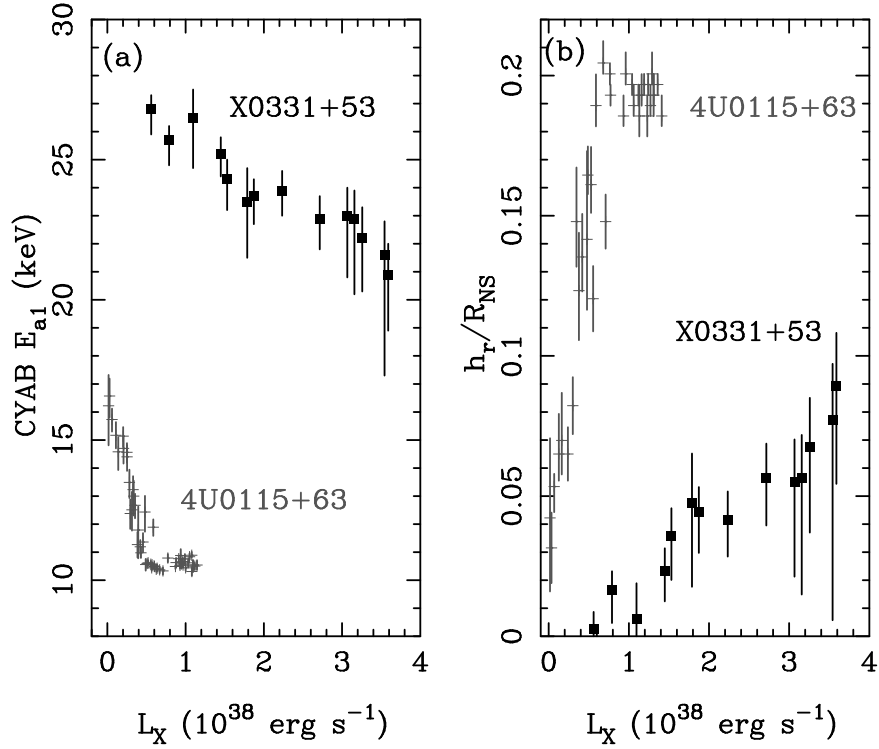}
\caption{Comparison between X0331+53 and 4U~0115+63, in terms of the luminosity dependence of the fundamental resonance energy (panel a) and of the estimated cyclotron ``photosphere'' height. The CRSF are modeled with eq.\ref{e2}. The results of 4U~0115+63 refer to Nakajima et al.\ (2006a).}
\label{f14}
\end{figure}

Let us fist compare the luminosity dependence of {\ea} in the two objects. 
As shown in Figure \ref{f14}a, 
the resonance energy of X0331+53 depends on 
the luminosity much less steeply than that of  4U~0115+63, 
even though the sense of dependence is the same between them. 
Specifically, the dependence is expressed as 
$(\Delta E_{a1}/E_{a1})/(\Delta L_{X}/L_{X}) \sim 0.08$ in X0331+53, 
while $(\Delta E_{a1}/E_{a1})/(\Delta L_{X}/L_{X}) \sim 0.26$ 
in 4U~0115+63.

\begin{figure}
\plotone{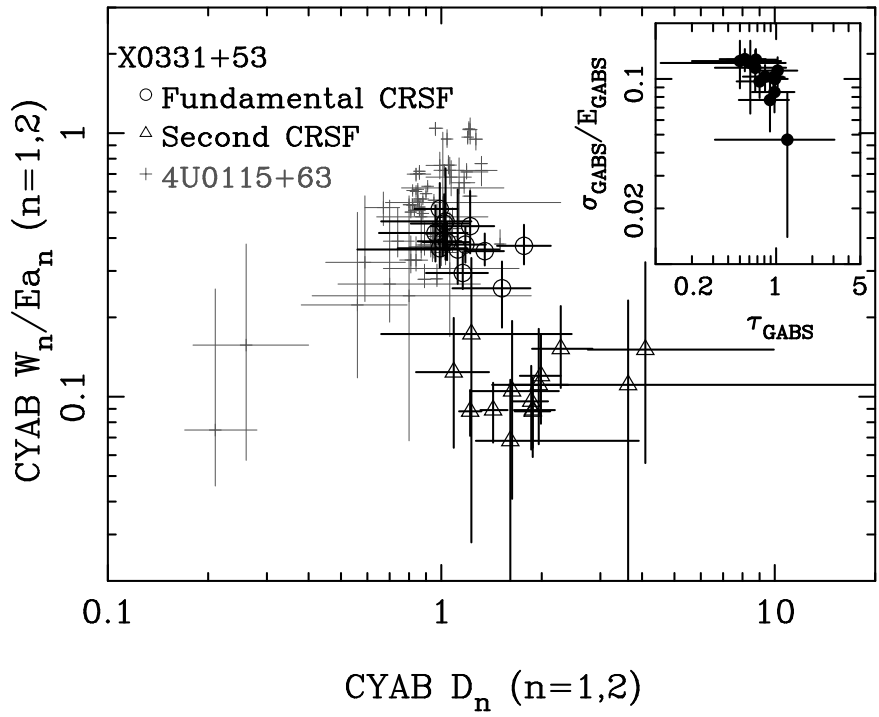}
\caption{CRSF depths $D$ of X0331+53, plotted against the fractional width $W/E_a$. Results on 4U~0115+63 \citep{m06a} are also shown for comparison. }
\label{f15}
\end{figure}

Assuming that the change in {\ea} is caused by 
variations in the height of cyclotron resonance ``photosphere'', 
the above difference suggests 
that the accretion column height of 4U~0115+63 
responds more sensitively to changes in the mass accretion rate. 
This inference is reinforced by Figure \ref{f14}b, 
which directly relates the estimated photosphere heights 
with the luminosity. 
This difference, in turn, may be attributed to 
differences in the accretion column shape. 
Indeed, as shown in Figure \ref{f15}, 
the resonance depths $D$ of X0331+53 depend 
negatively on the resonance width $W/E_{\rm a}$, 
while the correlation is opposite in 4U~0115+63. 
According to previous studies 
\citep[and references therein]{ise98,kre04,m06a}, 
this is considered to indicate 
that the accretion column in 4U~0115+63 has a tall cylindrical shape, 
whereas that in X0331+53 a flat coin-like shape.

The two sources are inferred to differ 
not only in the shape of the accretion column, 
but also in its area $A_{\rm col}$. 
We expect the observed luminosity to be expressed as 
$L_{\rm X} \cong A_{\rm col} \sigma_{{\rm SB}} T_{\rm eff}^4$ \citep{bil97}, 
where $\sigma_{{\rm SB}}$ is the Stefan-Boltzmann constant, 
and $T_{\rm eff}$ is the accretion column temperature 
which can be approximated by the observed NPEX $kT$ \citep{max99,tm04,m06a}. 
By substituting the derived parameters into this relation, 
we actually find X0331+53 to have a 1.43 times 
larger $A_{\rm col}$ than 4U~0115+63. 
The relatively shallow spin modulation in X0331+53 \citep{pot05} 
supports this idea. 
Therefore, the accretion column in X0331+53 is considered 
to have a coin-like shape and a rather large area, 
so that its height depends only weakly on the mass accretion rate.

Let us finally discuss the behavior of the resonance depths. 
As already mentioned in the end of section \ref{sec3.3}, 
$D_1$, $D_2$, and $\tau_{\rm ga}$ of X0331+53 have all 
been found to decrease toward higher luminosity. 
This change is in the same sense as those in 
4U~0115+63 \citep{m06a} and GX301-2 \citep{oka04}. 
One possible mechanism causing such a dependence 
may be related to the appearance of higher harmonic resonances. 
As argued by several authors \citep{m06,m06a}, 
an increased luminosity may give rise to 
significant higher harmonic resonances, 
which tend to make the fundamental resonance shallower 
through so-called ``two-photon'' effects \citep{alm91}. 
In fact, the third CRSF of X0331+53 was confirmed 
to appear when the source becomes luminous.

In summary, we have quantified luminosity dependence 
of the fundamental and second resonance energies, 
and discovered clear luminosity-dependent changes in their ratio. 
In addition, we have discovered that the behavior of X0331+53 
is qualitatively similar to that of 4U~0115+53, 
but differs quantitatively.


\clearpage

\begin{deluxetable}{l ccc cc l ccc}
\tabletypesize{\tiny}
\tablecaption{The log of {\it RXTE} observations of X0331+53 in the 2004$-$2005 outburst. \label{t1}}
\tablewidth{0pt}
\tablehead{
 \colhead{} & 
 \colhead{} & 
 \colhead{PCA} & 
 \colhead{HEXTE} &
 \colhead{} &  \colhead{} & 
 \colhead{} & 
 \colhead{} & 
 \colhead{PCA} & 
 \colhead{HEXTE} \\
 \colhead{Date} &  
 \colhead{Start/End Time\tablenotemark{a}} &
 \colhead{Rate\tablenotemark{b}} &
 \colhead{Rate\tablenotemark{c}} &
 \colhead{} & \colhead{} &
 \colhead{Date} &  
 \colhead{Start/End Time\tablenotemark{a}} &
 \colhead{Rate\tablenotemark{b}} &
 \colhead{Rate\tablenotemark{c}} \\
 \colhead{(2004/2005)} &
 \colhead{(UT)} &
 \colhead{[c s$^{-1}$]} &
 \colhead{[c s$^{-1}$]} &
 \colhead{} & \colhead{} &
 \colhead{(2004/2005)} &
 \colhead{(UT)} &
 \colhead{[c s$^{-1}$]} &
 \colhead{[c s$^{-1}$]} 
}
\startdata
Nov 27  & 14:06/14:13 &$  541\pm2 $&$  35.9\pm1.4 $& & & Jan 11d & 14:48/15:52 &$ 2049\pm4 $&$  73.3\pm0.4 $\\
Dec  1  & 06:54/07:41 &$  848\pm2 $&$  44.5\pm0.4 $& & & Jan 14  & 08:43/15:59 &$ 1841\pm4 $&$  66.8\pm0.2 $\\
Dec  2a & 03:20/04:12 &$  973\pm2 $&$  49.9\pm0.4 $& & & Jan 15a & 00:28/04:33 &$ 1789\pm4 $&$  64.9\pm0.2 $\\
Dec  2b & 19:16/20:11 &$ 1042\pm2 $&$  52.2\pm0.4 $& & & Jan 15b & 08:19/15:14 &$ 1770\pm3 $&$  64.4\pm0.2 $\\
Dec  3  & 18:52/00:25 &$ 1121\pm2 $&$  54.1\pm0.2 $& & & Jan 15c & 20:57/21:17 &$ 1708\pm4 $&$  61.7\pm0.5 $\\
Dec  4a & 01:00/01:30 &$ 1100\pm2 $&$  52.1\pm0.5 $& & & Jan 15d & 22:28/03:59 &$ 1707\pm3 $&$  61.3\pm0.2 $\\
Dec  4b & 02:34/03:14 &$ 1141\pm2 $&$  55.2\pm0.4 $& & & Jan 16a & 09:30/14:29 &$ 1623\pm3 $&$  60.7\pm0.2 $\\
Dec  4c & 05:52/06:25 &$ 1155\pm2 $&$  54.3\pm0.6 $& & & Jan 16b & 20:33/02:12 &$ 1604\pm3 $&$  59.5\pm0.2 $\\
Dec  4d & 09:15/09:39 &$ 1189\pm2 $&$  57.2\pm0.5 $& & & Jan 17a & 02:49/03:36 &$ 1578\pm3 $&$  59.2\pm0.4 $\\
Dec  4e & 12:00/18:02 &$ 1176\pm2 $&$  56.6\pm0.2 $& & & Jan 17b & 07:31/13:00 &$ 1541\pm3 $&$  58.1\pm0.2 $\\
Dec  4f & 18:21/00:00 &$ 1204\pm2 $&$  56.3\pm0.2 $& & & Jan 17c & 20:10/00:13 &$ 1508\pm3 $&$  56.8\pm0.2 $\\
Dec  5a & 00:36/01:08 &$ 1219\pm2 $&$  56.4\pm0.5 $& & & Jan 18a & 00:50/01:55 &$ 1475\pm3 $&$  56.1\pm0.3 $\\
Dec  5b & 13:11/14:16 &$ 1287\pm3 $&$  60.0\pm0.4 $& & & Jan 18b & 07:08/12:58 &$ 1316\pm3 $&$  56.0\pm0.2 $\\
Dec  5c & 17:55/19:15 &$ 1328\pm3 $&$  60.7\pm0.5 $& & & Jan 18c & 21:17/01:24 &$ 1433\pm3 $&$  54.8\pm0.2 $\\
Dec  6  & 17:36/18:51 &$ 1443\pm3 $&$  65.7\pm0.5 $& & & Jan 19a & 02:01/02:46 &$ 1440\pm3 $&$  55.0\pm0.4 $\\
Dec  7  & 17:05/17:51 &$ 1545\pm3 $&$  69.7\pm0.5 $& & & Jan 19b & 03:39/08:41 &$ 1424\pm3 $&$  54.0\pm0.2 $\\
Dec 13  & 18:30/19:11 &$ 2392\pm5 $&$  88.9\pm0.6 $& & & Jan 20  & 07:53/11:30 &$ 1366\pm3 $&$  54.9\pm0.2 $\\
Dec 14  & 17:25/18:21 &$ 2557\pm5 $&$  93.7\pm0.4 $& & & Jan 21  & 07:42/08:21 &$ 1295\pm3 $&$  53.1\pm0.4 $\\
Dec 15a & 12:18/14:20 &$ 2609\pm5 $&$  95.7\pm0.3 $& & & Jan 23  & 05:07/06:32 &$ 1201\pm2 $&$  50.3\pm0.3 $\\
Dec 15b & 14:51/19:42 &$ 2614\pm5 $&$  96.9\pm0.2 $& & & Jan 24a & 06:23/13:49 &$ 1183\pm2 $&$  49.5\pm0.2 $\\
Dec 15c & 20:36/21:17 &$ 2604\pm5 $&$  95.4\pm0.5 $& & & Jan 24b & 14:13/15:07 &$ 1165\pm2 $&$  48.9\pm0.4 $\\
Dec 16  & 02:26/03:02 &$ 2665\pm5 $&$  96.2\pm0.5 $& & & Jan 25  & 07:26/10:36 &$ 1099\pm2 $&$  48.4\pm0.2 $\\
Dec 17  & 11:44/12:51 &$ 2789\pm6 $&$  97.2\pm0.4 $& & & Jan 28  & 10:57/11:43 &$  943\pm2 $&$  43.7\pm0.4 $\\
Dec 18  & 01:41/02:53 &$ 2851\pm6 $&$ 100.5\pm0.5 $& & & Jan 29  & 14:25/14:51 &$  927\pm2 $&$  43.9\pm0.6 $\\
Dec 19a & 14:14/15:10 &$ 2870\pm6 $&$ 100.9\pm0.4 $& & & Jan 31  & 08:48/09:13 &$  858\pm2 $&$  41.8\pm0.5 $\\
Dec 19b & 20:08/20:32 &$ 2859\pm6 $&$  98.1\pm0.7 $& & & Feb  2  & 08:09/08:23 &$  780\pm2 $&$  37.8\pm0.7 $\\
Dec 20  & 23:58/00:58 &$ 3017\pm6 $&$ 105.4\pm0.5 $& & & Feb  4  & 05:39/05:54 &$  780\pm2 $&$  38.5\pm0.6 $\\
Dec 22  & 02:49/03:32 &$ 3023\pm6 $&$ 103.9\pm0.7 $& & & Feb  6a & 13:34/16:19 &$  681\pm1 $&$  35.2\pm0.2 $\\
Dec 24  & 09:20/16:11 &$ 3163\pm6 $&$ 101.7\pm0.2 $& & & Feb  6b & 18:16/18:45 &$  696\pm1 $&$  35.8\pm0.4 $\\
Dec 25  & 08:32/15:48 &$ 3120\pm6 $&$ 100.0\pm0.1 $& & & Feb  6c & 23:00/23:36 &$  665\pm1 $&$  35.7\pm0.4 $\\
Dec 28  & 05:28/06:21 &$ 3119\pm6 $&$  99.0\pm0.5 $& & & Feb  8  & 06:46/07:34 &$  618\pm1 $&$  32.9\pm0.3 $\\
Dec 29a & 05:52/06:30 &$ 3078\pm6 $&$  99.5\pm0.4 $& & & Feb  9  & 18:02/18:14 &$  545\pm2 $&$  30.8\pm1.0 $\\
Dec 29b & 21:52/22:29 &$ 3098\pm6 $&$ 102.4\pm0.6 $& & & Feb 10  & 14:52/15:26 &$  516\pm1 $&$  29.6\pm0.5 $\\
Dec 30  & 13:35/14:20 &$ 3040\pm6 $&$  97.7\pm0.5 $& & & Feb 12a & 01:23/04:10 &$  478\pm1 $&$  24.2\pm1.1 $\\
Jan  6a & 06:23/06:38 &$ 2582\pm5 $&$  85.5\pm0.8 $& & & Feb 12b & 17:04/21:06 &$  446\pm1 $&$  24.4\pm0.3 $\\
Jan  6b & 15:15/16:04 &$ 2626\pm5 $&$  87.4\pm0.4 $& & & Feb 13a & 00:58/01:45 &$  440\pm1 $&$  25.6\pm0.6 $\\
Jan  8a & 10:11/10:34 &$ 2475\pm5 $&$  79.0\pm0.6 $& & & Feb 13b & 04:12/08:18 &$  427\pm1 $&$  25.9\pm0.2 $\\
Jan  8b & 14:25/14:56 &$ 2490\pm5 $&$  83.7\pm0.6 $& & & Feb 13c & 11:45/14:10 &$  401\pm1 $&$  24.4\pm0.2 $\\
Jan  8c & 17:41/18:25 &$ 2445\pm5 $&$  83.3\pm0.5 $& & & Feb 13d & 18:01/19:01 &$  390\pm1 $&$  23.6\pm0.3 $\\
Jan 10  & 10:57/16:02 &$ 2157\pm4 $&$  76.5\pm0.2 $& & & Feb 15  & 12:07/14:56 &$  329\pm1 $&$  21.0\pm0.3 $\\
Jan 11a & 01:01/01:24 &$ 2112\pm4 $&$  72.7\pm0.5 $& & & Mar  7  & 07:07/08:06 &$ 33.8\pm0.1$&$  2.6\pm0.2 $\\
Jan 11b & 08:20/09:18 &$ 2159\pm4 $&$  74.2\pm0.4 $& & & Mar  9  & 07:54/08:10 &$ 75.9\pm0.5$&$  5.1\pm0.7 $\\
Jan 11c & 11:32/12:32 &$ 2106\pm4 $&$  74.0\pm0.4 $& & & Mar 18  & 01:34/01:50 &$ 12.4\pm0.2$&$  0.8\pm0.4 $\\
\hline
\enddata

\tablenotetext{a}{
Start and end time (UT) of the PCA observations. 
}
\tablenotetext{b}{
Background subtracted count rates of PCU2 in the 3-20 keV energy range. 
}
\tablenotetext{c}{
Background subtracted count rates of HEXTE cluster B in the 20-80 keV energy range.
}
\end{deluxetable}

\clearpage

\begin{deluxetable}{lc cccc}
\tablecaption{The best-fit parameters of NPEX$\times$CYAB2$\times$GABS model, determined by the four representative spectra shown in Figure \ref{f2}. \label{t2}}
\tablewidth{0pt}
\tablehead{
 \colhead{} & \colhead{} & 
 \multicolumn{4}{c}{Obs. Date} \\
 \cline{2-6} \\
 \colhead{parameters} & 
 \colhead{} &  
 \colhead{Dec 2a} &   
 \colhead{Dec 24} &   
 \colhead{Jan 20} &   
 \colhead{Feb 13b} \\ 
}
\startdata
$I_{iron}^a$    & &$ 1.54\pm0.26           $&$ 6.64_{-0.99}^{+1.03}  $&$ 2.59_{-0.35}^{+0.36}  $&$ 0.65_{-0.08}^{+0.11} $\\
$\alpha_1$      & &$ -0.22\pm0.07          $&$ -0.22_{-0.10}^{+0.13} $&$ -0.30_{-0.08}^{+0.10} $&$ 0.07_{-0.08}^{+0.09} $\\
$kT$ (keV)      & &$ 6.25_{-1.18}^{+2.82}  $&$ 5.62_{-0.16}^{+0.17}  $&$ 5.57_{-0.79}^{+0.90}  $&$ 6.16_{-0.61}^{+0.78} $\\
$E_{a1}$ (keV)  & &$ 25.8_{-1.2}^{+0.6}    $&$ 21.1_{-2.1}^{+1.0}    $&$ 24.4_{-1.5}^{+0.9}    $&$ 25.4_{-1.2}^{+1.0}   $\\
$W_{1}$ (keV)   & &$ 8.69_{-1.86}^{+1.60}  $&$ 10.71_{-1.67}^{+2.96} $&$ 8.61_{-1.44}^{+2.00}  $&$ 9.83_{-1.52}^{+1.11} $\\
$D_{1}$         & &$ 1.32_{-0.29}^{+0.27}  $&$ 1.10_{-0.19}^{+0.11}  $&$ 1.14_{-0.36}^{+0.24}  $&$ 1.05_{-0.27}^{+0.35} $\\
$E_{a2}$ (keV)  & &$ 51.7~({\rm fixed})    $&$ 49.5\pm0.3            $&$ 51.5_{-1.6}^{+1.3}    $&$ 50.8~({\rm fixed})   $\\
$W_{2}$ (keV)   & &$ 7.78_{-3.04}^{+6.19}  $&$ 4.86_{-1.07}^{+1.17}  $&$ 5.60_{-4.60}^{+6.22}  $&$ 9.83 ({\rm fixed})   $\\
$D_{2}$         & &$ 1.77_{-0.90}^{+0.75}  $&$ 1.35\pm0.12           $&$ 1.95_{-0.61}^{+6.99}  $&$ 1.90_{-0.30}^{+0.47} $\\
$E_{\rm ga}$ (keV)
                & &$ 29.8_{-0.7}^{+1.0}    $&$ 26.4\pm0.3            $&$ 29.3_{-0.4}^{+0.5}    $&$ 29.7_{-0.3}^{+0.5}   $\\   
$\sigma_{\rm ga}$ (keV)							 
                & &$ 2.41_{-1.35}^{+0.73}  $&$ 3.14_{-0.57}^{+0.63}  $&$ 2.70_{-0.91}^{+0.72}  $&$ 2.42_{-0.43}^{+0.56} $\\
$\tau_{\rm ga}$ & &$ 0.90_{-0.63}^{+0.55}  $&$ 0.48_{-0.20}^{+0.34}  $&$ 0.92_{-0.46}^{+0.58}  $&$ 1.33_{-0.47}^{+0.46} $\\
$L_{\rm X}$ $^b$& &$  1.37                 $&$  3.65                 $&$  1.76                 $&$  0.65                $\\
$\chi_{\nu}^{2}(\nu)$
                & &$  0.77(52)             $&$  0.55(51)             $&$  0.66(51)             $&$  0.67(52)            $\\
\hline
\enddata

\tablenotetext{a}{
In units of $10^{-2}$ photons cm$^{-1}$ s$^{-1}$. 
}
\tablenotetext{b}{
In units of $\times10^{38}$ ergs s$^{-1}$ in $3-80$ keV. 
}
\end{deluxetable}

\clearpage

\begin{deluxetable}{lcccccccccccccc ccccccc}
\tabletypesize{\tiny}
\rotate
\tablecaption{The best-fit NPEX$\times$CYAB2$\times$GABS model parameters, determined by the flux-sorted spectra. \label{t3}}
\tablewidth{0pt}
\tablehead{
 \colhead{} &
 \multicolumn{2}{c}{NPEX} &
 \colhead{} & 
 \multicolumn{6}{c}{CYAB} & 
 \colhead{} & 
 \multicolumn{3}{c}{GABS} & 
 \colhead{} &
 \colhead{} \\ \cline{2-3} \cline{5-10} \cline{12-14} 
 \colhead{flux} & 
 \colhead{$\alpha_1$} & 
 \colhead{$kT$} & 
 \colhead{} &
 \colhead{$E_{a1}$} & 
 \colhead{$W_1$} & 
 \colhead{$D_1$} & 
 \colhead{$E_{a2}$} & 
 \colhead{$W_2$} & 
 \colhead{$D_2$} & 
 \colhead{} &
 \colhead{$E_{\rm ga}$} & 
 \colhead{$\sigma_{\rm ga}$} & 
 \colhead{$\tau_{\rm ga}$} & 
 \colhead{${\rm L_X}$\tablenotemark{a}} &
 \colhead{$\chi_\nu^2$} \\
 \colhead{level} & 
 \colhead{} &  
 \colhead{(keV)} &
 \colhead{} &  
 \colhead{(keV)} &
 \colhead{(keV)} &
 \colhead{} &
 \colhead{(keV)} &
 \colhead{(keV)} &
 \colhead{} &
 \colhead{} &
 \colhead{(keV)} &
 \colhead{(keV)} &
 \colhead{} &
 \colhead{} &
 \colhead{} \\
}
\startdata
\multicolumn{15}{l}{ascent phase} \\
%
%
fa1 &$ -0.23_{-0.11}^{+0.13} $&$ 5.43_{-0.12}^{+0.14} $& &$ 20.9_{-2.0}^{+1.1} $&$ 10.79_{-1.83}^{+2.68}$&$ 0.99_{-0.16}^{+0.13} $&$ 49.3\pm0.3         $&$ 4.33_{-0.86}^{+0.88} $&$ 1.22_{-0.09}^{+0.10} $& &$ 26.4_{-0.3}^{+0.4} $&$ 3.39_{-0.48}^{+0.46} $&$ 0.55_{-0.21}^{+0.27} $& 3.59 & 0.81 \\
fa2 &$ -0.24_{-0.10}^{+0.12} $&$ 5.46_{-0.15}^{+0.17} $& &$ 22.2_{-1.9}^{+1.1} $&$ 10.08_{-1.85}^{+2.92}$&$ 1.02_{-0.21}^{+0.20} $&$ 49.5\pm0.3         $&$ 4.40_{-1.10}^{+1.18} $&$ 1.43_{-0.12}^{+0.15} $& &$ 27.4_{-0.3}^{+0.5} $&$ 3.47_{-0.56}^{+0.49} $&$ 0.68_{-0.29}^{+0.34} $& 3.26 & 0.59 \\
fa3 &$ -0.25_{-0.09}^{+0.12} $&$ 6.03_{-0.39}^{+0.50} $& &$ 23.0_{-2.2}^{+1.0} $&$ 10.21_{-1.93}^{+3.62}$&$ 1.22_{-0.29}^{+0.22} $&$ 50.3_{-0.6}^{+0.5} $&$ 6.05_{-2.08}^{+2.59} $&$ 1.99_{-0.27}^{+0.29} $& &$ 27.6_{-0.3}^{+0.5} $&$ 3.17_{-0.83}^{+0.78} $&$ 0.67_{-0.36}^{+0.54} $& 3.07 & 0.71 \\
fa5 &$ -0.24_{-0.10}^{+0.13} $&$ 5.58_{-0.50}^{+0.71} $& &$ 23.5_{-2.0}^{+1.2} $&$ 9.83_{-2.00}^{+2.64} $&$ 0.96_{-0.31}^{+0.24} $&$ 51.1_{-1.0}^{+1.1} $&$ 5.35_{-3.27}^{+4.55} $&$ 1.63_{-0.41}^{+0.62} $& &$ 29.1_{-0.3}^{+0.4} $&$ 3.23_{-0.58}^{+0.59} $&$ 1.03_{-0.36}^{+0.47} $& 1.79 & 0.57 \\
fa6 &$ -0.20_{-0.09}^{+0.10} $&$ 5.60_{-0.23}^{+0.25} $& &$ 24.3_{-1.1}^{+0.7} $&$ 9.37_{-1.29}^{+1.74} $&$ 1.04_{-0.16}^{+0.18} $&$ 51.1\pm0.4         $&$ 4.53_{-1.29}^{+1.47} $&$ 1.88_{-0.23}^{+0.31} $& &$ 29.3_{-0.2}^{+0.3} $&$ 2.93_{-0.41}^{+0.34} $&$ 0.98\pm0.27          $& 1.53 & 0.55 \\
fa7 &$ -0.25_{-0.06}^{+0.11} $&$ 6.52_{-1.36}^{+5.14} $& &$ 26.5_{-1.8}^{+1.0} $&$ 6.84_{-1.93}^{+1.79} $&$ 1.52_{-0.44}^{+0.33} $&$ 51.8_{-2.9}^{+8.1} $&$ 5.75_{-4.75}^{+6.22} $&$ 3.63_{-2.21}^{+60.76}$& &$ 31.3_{-1.3}^{+1.1} $&$ 0.91_{-0.42}^{+0.60} $&$ 9.15({\rm fixed})    $& 1.10 & 0.64 \\
\multicolumn{15}{c}{} \\
\multicolumn{15}{l}{descent phase} \\
fd1 &$ -0.29_{-0.10}^{+0.13} $&$ 5.56_{-0.36}^{+0.51} $& &$ 21.6_{-4.3}^{+1.2} $&$ 10.00_{-2.04}^{+5.88}$&$ 1.03_{-0.37}^{+0.20} $&$ 49.3\pm0.9         $&$ 6.11_{-2.95}^{+3.70} $&$ 1.09_{-0.25}^{+0.30} $& &$ 26.5_{-0.6}^{+0.7} $&$ 3.32_{-0.95}^{+0.95} $&$ 0.50_{-0.30}^{+0.60} $& 3.54 & 0.64 \\
fd2 &$ -0.29_{-0.09}^{+0.12} $&$ 5.52_{-0.72}^{+1.84} $& &$ 22.9_{-2.7}^{+1.0} $&$ 8.30_{-1.91}^{+5.70} $&$ 1.12_{-0.56}^{+0.37} $&$ 49.7_{-1.7}^{+1.6} $&$ 8.61_{-7.19}^{+8.08} $&$ 1.23_{-0.57}^{+1.23} $& &$ 27.8_{-1.0}^{+1.1} $&$ 3.38_{-1.57}^{+1.07} $&$ 0.61_{-0.50}^{+0.59} $& 3.16 & 0.53 \\
fd3 &$ -0.24_{-0.09}^{+0.10} $&$ 5.09_{-0.28}^{+0.32} $& &$ 22.9_{-1.1}^{+0.8} $&$ 8.37_{-1.25}^{+1.43} $&$ 0.99_{-0.25}^{+0.21} $&$ 50.3_{-0.8}^{+0.7} $&$ 3.40_{-2.40}^{+2.39} $&$ 1.61_{-0.34}^{+2.30} $& &$ 28.3_{-0.3}^{+0.4} $&$ 2.92_{-0.51}^{+0.45} $&$ 0.81_{-0.28}^{+0.33} $& 2.72 & 0.80 \\
fd4 &$ -0.26_{-0.08}^{+0.10} $&$ 5.37_{-0.27}^{+0.32} $& &$ 23.9_{-0.9}^{+0.7} $&$ 9.03_{-1.28}^{+1.48} $&$ 1.18_{-0.17}^{+0.18} $&$ 51.7\pm0.5         $&$ 4.94_{-1.68}^{+1.79} $&$ 1.86_{-0.23}^{+0.23} $& &$ 28.6_{-0.2}^{+0.3} $&$ 2.77_{-0.51}^{+0.38} $&$ 0.73\pm0.26          $& 2.23 & 0.61 \\
fd5 &$ -0.26_{-0.08}^{+0.11} $&$ 5.31_{-0.25}^{+0.23} $& &$ 23.7_{-1.0}^{+0.6} $&$ 9.23_{-1.05}^{+1.47} $&$ 1.01_{-0.16}^{+0.14} $&$ 51.6_{-0.5}^{+0.4} $&$ 4.56_{-1.52}^{+1.33} $&$ 1.88_{-0.22}^{+0.24} $& &$ 28.9\pm0.2         $&$ 2.94\pm0.28          $&$ 0.95_{-0.19}^{+0.23} $& 1.87 & 0.68 \\
fd6 &$ -0.31_{-0.06}^{+0.07} $&$ 6.28_{-0.55}^{+1.04} $& &$ 25.2_{-0.8}^{+0.6} $&$ 8.99_{-1.01}^{+1.47} $&$ 1.35_{-0.18}^{+0.19} $&$ 52.4_{-0.7}^{+1.0} $&$ 7.95_{-2.29}^{+3.62} $&$ 2.28_{-0.41}^{+0.56} $& &$ 29.3\pm0.3         $&$ 2.27_{-0.74}^{+0.55} $&$ 0.89_{-0.40}^{+0.39} $& 1.45 & 0.44 \\
fd7 &$ -0.04_{-0.10}^{+0.09} $&$ 5.54_{-0.37}^{+0.70} $& &$ 25.7_{-0.9}^{+0.5} $&$ 7.58_{-0.97}^{+1.52} $&$ 1.16_{-0.26}^{+0.22} $&$ 51.8_{-0.7}^{+1.0} $&$ 5.74_{-2.32}^{+3.62} $&$ 1.96_{-0.37}^{+0.44} $& &$ 30.1_{-0.4}^{+0.3} $&$ 2.56_{-0.58}^{+0.55} $&$ 0.97_{-0.34}^{+0.45} $& 0.79 & 1.08 \\
fd8 &$  0.07_{-0.06}^{+0.09} $&$ 7.24_{-1.22}^{+8.88} $& &$ 26.8_{-0.9}^{+0.5} $&$ 10.03_{-1.46}^{+2.01}$&$ 1.77_{-0.30}^{+0.36} $&$ 54.4_{-2.0}^{+4.1} $&$ 8.24_{-5.15}^{+9.40} $&$ 4.09_{-1.34}^{+5.84} $& &$ 29.2\pm0.7         $&$ 1.38_{-0.97}^{+1.12} $&$ 1.24_{-0.93}^{+1.81} $& 0.56 & 0.42 \\
\hline
\enddata

\tablenotetext{a}
{
$\times10^{38}$ ergs s$^{-1}$ in $3-80$ keV
}

\end{deluxetable}

\begin{deluxetable}{lcccccccccccccc ccccccc}
\tabletypesize{\tiny}
\rotate
\tablecaption{The best-fit PLCUT$\times$CYAB2$\times$GABS model parameters, determined by the flux-sorted spectra.\label{t4}}
\tablewidth{0pt}
\tablehead{
 \colhead{} &
 \multicolumn{3}{c}{PLCUT} &
 \colhead{} & 
 \multicolumn{6}{c}{CYAB} & 
 \colhead{} & 
 \multicolumn{3}{c}{GABS} & 
 \colhead{} &
 \colhead{} \\ \cline{2-4} \cline{6-11} \cline{13-15} 
 \colhead{flux} & 
 \colhead{$\Gamma$} & 
 \colhead{$E_{\rm cut}$} & 
 \colhead{$E_{\rm fold}$} & 
 \colhead{} &
 \colhead{$E_{a1}$} & 
 \colhead{$W_1$} & 
 \colhead{$D_1$} & 
 \colhead{$E_{a2}$} & 
 \colhead{$W_2$} & 
 \colhead{$D_2$} & 
 \colhead{} &
 \colhead{$E_{\rm ga}$} & 
 \colhead{$\sigma_{\rm ga}$} & 
 \colhead{$\tau_{\rm ga}$} & 
 \colhead{${\rm L_X}$\tablenotemark{a}} &
 \colhead{$\chi_\nu^2$} \\
 \colhead{level} & 
 \colhead{} &  
 \colhead{(keV)} &
 \colhead{(keV)} &
 \colhead{} &  
 \colhead{(keV)} &
 \colhead{(keV)} &
 \colhead{} &
 \colhead{(keV)} &
 \colhead{(keV)} &
 \colhead{} &
 \colhead{} &
 \colhead{(keV)} &
 \colhead{(keV)} &
 \colhead{} &
 \colhead{} &
 \colhead{} \\
}
\startdata
fd1  &$ -0.34_{-0.07}^{+0.09} $&$  2.24_{-1.24}^{+1.24} $&$  7.19_{-0.53}^{+0.73} $& &$ 22.5_{-3.1}^{+0.6} $&$ 7.75_{-1.03}^{+8.59} $&$ 1.03_{-0.70}^{+0.17} $&$ 49.8_{-0.9}^{+0.9} $&$ 7.68_{-2.79}^{+4.10} $&$ 1.17_{3.71}^{+10.61} $& &$ 27.24_{-1.97}^{+0.78} $&$ 3.20_{-1.01}^{+1.47} $&$ 0.40_{-0.24}^{+1.08} $& 35.42 & 0.74 \\
fd6  &$ -0.26_{-0.07}^{+0.14} $&$  1.69_{-0.69}^{+1.95} $&$ 10.16_{-0.80}^{+2.13} $& &$ 25.3_{-1.0}^{+0.5} $&$ 8.44_{-0.69}^{+1.22} $&$ 1.38_{-0.22}^{+0.17} $&$ 53.5_{-0.8}^{+1.0} $&$ 12.64_{-2.10}^{+3.76}$&$  2.96_{7.58}^{+13.44} $& &$ 29.55_{-0.32}^{+0.32} $&$ 2.33_{-0.76}^{+0.63} $&$ 0.87_{-0.39}^{+0.50} $& 14.50 & 0.61 \\
\hline
\enddata

\tablenotetext{a}
{
$\times10^{38}$ ergs s$^{-1}$ in $3-80$ keV
}

\end{deluxetable}


\clearpage

\clearpage

\clearpage

\clearpage

\clearpage

\clearpage

\clearpage

\clearpage

\clearpage

\clearpage

\clearpage

\clearpage

\clearpage

\clearpage

\clearpage

\end{document}